\documentclass[onecolumn, aps, nofootinbib, prd, preprint, floats, floatfix, amsmath, amssymb, superscriptaddress, preprintnumbers]{revtex4-1}
\usepackage{slashed}
\usepackage{graphicx}
\usepackage{subcaption}
\usepackage{dcolumn}
\usepackage{bm}
\usepackage{amssymb}
\usepackage[utf8]{inputenc}
\usepackage{yfonts,amsmath,amsthm,amsfonts,amssymb,amscd, ulem}
\usepackage{enumerate}
\usepackage{fancyhdr}
\usepackage{mathtools}
\usepackage{mathrsfs}
\usepackage{cancel}
\usepackage{slashed}
\usepackage{bigints}
\usepackage[flushleft]{threeparttable}
\usepackage[colorlinks=true, citecolor=purple, linkcolor=blue]{hyperref}
\usepackage{url}
\usepackage{makecell,booktabs}
\usepackage{braket}
\usepackage{relsize}
\usepackage{multirow}
\usepackage{verbatim}
\usepackage{txfonts}
\usepackage{slashed}
\usepackage{upgreek}
\usepackage{extarrows}
\usepackage{array}
\usepackage{appendix}
\usepackage[T1]{fontenc}
\usepackage{setspace}
\usepackage{subcaption}

\usepackage[dvipsnames]{xcolor}
\definecolor{c1}{HTML}{1bd1a5}

\usepackage{graphicx}   
\usepackage{caption}    
\usepackage{subcaption} 
\usepackage{subcaption}
\usepackage{caption}
\usepackage{tikz} 
\usetikzlibrary{shapes,arrows,positioning,automata,backgrounds,calc,er,patterns}
\usepackage{tikz-feynman}
\tikzfeynmanset{compat=1.0.0}
\usetikzlibrary{shapes.misc}
\tikzset{cross/.style={cross out, draw=black, fill=none, minimum size=2*(#1-\pgflinewidth), inner sep=0pt, outer sep=0pt}, cross/.default={2pt}}
\usetikzlibrary{shapes.geometric}
\usepackage{orcidlink}

\begin{document}

\title{Ultralight Scalar Dark Matter with Off-Diagonal Flavor Couplings}

\author{Jinhui Guo}
\email{guojh23@buaa.edu.cn}
\affiliation{School of Physics, Beihang University, Beijing 100191, China}

\author{Jia Liu \orcidlink{0000-0001-7386-0253}}
\email{jialiu@pku.edu.cn}
\affiliation{School of Physics and State Key Laboratory of Nuclear Physics and Technology, Peking University, Beijing 100871, China}
\affiliation{Center for High Energy Physics, Peking University, Beijing 100871, China}

\author{Chenhao Peng}
\email{chenhaopeng@stu.pku.edu.cn}
\affiliation{School of Physics and State Key Laboratory of Nuclear Physics and Technology, Peking University, Beijing 100871, China}

\author{Xiao-Ping Wang \orcidlink{0000-0002-2258-7741} }
\email{hcwangxiaoping@buaa.edu.cn}
\affiliation{School of Physics, Beihang University, Beijing 100191, China}

\author{Hang Zhao}
\email{hangzhao@buaa.edu.cn}
\affiliation{School of Physics, Beihang University, Beijing 100191, China}

\preprint{$\begin{gathered}\includegraphics[width=0.05\textwidth]{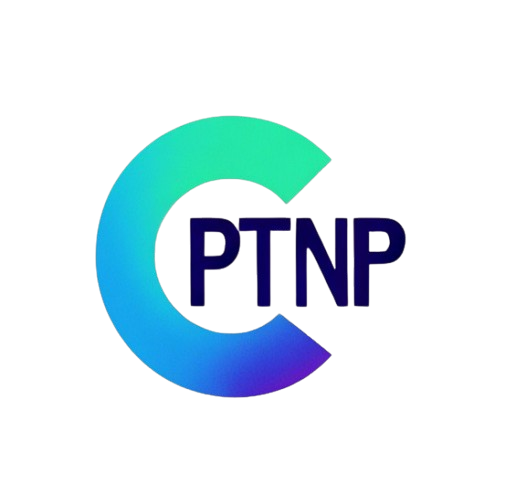}\end{gathered}$\, CPTNP-2026-013}

\begin{abstract}
Ultralight dark matter can behave as a coherent background field and induce time-dependent modifications of Standard Model parameters. We study a scenario in which a real ultralight scalar $\phi$ couples off-diagonally to down-type quarks, linking ultralight dark sectors to flavor physics.
Working within an effective field theory, we diagonalize the quark mass matrix in a coherent $\phi$ background and derive analytic expressions for oscillatory shifts in down-type quark masses and CKM parameters. These effects lead to signatures in both the classical regime, where $\phi$ acts as a background field, and the quantum (particle) regime, where it contributes through on-shell production or off-shell mediation.
Using precision flavor measurements, nuclear $\beta$ decays, atomic clocks, pulsar timing, and meson observables, we derive constraints on the flavor-violating couplings $\lambda_{ij}$ for $m_\phi\sim10^{-24}$–$10^{-12},\mathrm{eV}$, highlighting the complementarity of time-domain and flavor probes of ultralight dark sectors.
\end{abstract}

\maketitle

\tableofcontents

\section{Introduction}
\label{sec:int}
Dark matter accounts for about 85\% of the matter content of the Universe, yet its microscopic nature remains a central open problem in particle physics and cosmology \cite{Planck:2018vyg,Rubin:1970zza}. Ultralight dark matter (ULDM), often modeled as a weakly coupled oscillating real scalar field with masses spanning $m_\phi\sim10^{-24}\,{\rm eV}$ to ${\cal O}({\rm eV})$, is a well-motivated framework \cite{Ferreira:2020fam}. Owing to its macroscopic de Broglie wavelength, ULDM can behave coherently on astrophysical scales and is probed most efficiently through precision measurements rather than conventional heavy-dark-matter searches. In particular, even feeble couplings of ULDM to Standard Model (SM) fields can induce time-dependent variations in high-precision observables, such as atomic transition frequencies and nuclear decay rates, providing sensitive probes of ULDM interactions with the SM \cite{Uzan:2002vq,Safronova:2017xyt,Battaglieri:2017aum,Antypas:2022asj,Uzan:2010pm,Marsh:2021jmi,Arvanitaki:2014faa,Stadnik:2014tta,Stadnik:2015kia,Pospelov:2012mt,Derevianko:2013oaa,Stadnik:2014cea}.

Existing literature on scalar ULDM predominantly explores flavor-diagonal portals—often termed dilaton-like couplings—where $\phi$ interacts with SM fermion mass terms. These interactions manifest as oscillations in fermion masses and related fundamental constants \cite{SevillanoMunoz:2024ayh,Sherrill:2023zah,Hees:2016gop,Damour:2010rp,BACON:2020ubh}, and are stringently constrained by atomic-clock comparisons, interferometry, and tests of the equivalence principle \cite{Elder:2025tue,Broggini:2024udi,Dmitriev:2002kv,Gottel:2024cfj,Hall:2022zvi,Thirolf:2019ocm,Thirolf:2024xlx}. For a representative scalar interaction $\mathcal{L}_{\rm int}\sim \phi \sum_{f} \frac{m_f}{M_f}\bar f f$, the differential nucleon-electron scale $M_{\rm eff}\equiv (M^{-1}-M_e^{-1})^{-1}$ is already pushed to $M_{\rm eff}/M_{\rm Pl}\gtrsim 10^{5}$ (taking $M_p=M_n\equiv M$) across much of the ULDM mass range \cite{Elder:2025tue}. In contrast, the quark flavor sector remains relatively unexplored. In generic UV completions, couplings defined in the weak basis are not guaranteed to remain diagonal following electroweak symmetry breaking and the subsequent rotation to the mass basis. Consequently, off-diagonal $\phi$-quark interactions arise naturally, necessitating a systematic investigation of flavor-violating ULDM portals. This work focuses on the down-type sector, where such interactions directly modulate the Cabibbo-Kobayashi-Maskawa (CKM) matrix structure, subjecting ULDM to a diverse array of laboratory constraints.

We begin with an effective Lagrangian containing off-diagonal ULDM couplings to down-type quarks. In Sec.~\ref{sec:model}, we demonstrate that a coherent ULDM background induces three key effects: (i) shifts in down-type quark masses, (ii) time-dependent modifications of the CKM matrix, and (iii) derivative-induced kinetic mixings. We also provide the complementary particle interpretation relevant for decay and force searches. 
In Sec.~\ref{sec:uldm}, by treating $\phi$ as a classical condensate, we translate these effects into constraints derived from direct CKM determinations, time-resolved nuclear $\beta$-decay data, tritium decay measurements, atomic-clock comparisons, pulsar timing and meson mixing. 
Sec.~\ref{sec:particle} considers $\phi$ as an on-shell particle or an off-shell mediator, deriving bounds from quark-flavor-violating meson decays, meson oscillations, and loop-induced fifth-force effects. Additional astrophysical considerations are summarized in Sec.~\ref{sec:others}, and we conclude in Sec.~\ref{sec:concl}.

\section{A Flavor-violating ULDM Model}
\label{sec:model}
In this section, we present a framework for studying flavor-violating interactions between ULDM and SM quarks. We concentrate on a real scalar ULDM field, represented by $\phi$, which connects to quark generations through off-diagonal interactions. Due to its extremely small mass, ULDM can be described both as a coherent classical field and as particles. We develop these two perspectives in parallel and demonstrate that they yield identical physical consequences.

The effective Lagrangian in the SM quark mass eigenbasis is given by
\begin{equation}
    \mathcal{L}\supset \frac{1}{2}\partial_\mu\phi\partial^\mu\phi-\frac{1}{2}m_\phi^2\phi^2-\phi \sum_{i=1}^3\sum_{j=1,\,j\neq i}^3 \lambda_{ij} \bar{d}_i d_j,
\label{eq:eff_lag}
\end{equation}
where $m_\phi$ represents the mass of the ULDM field $\phi$, $d_i$ denotes the $i$-th generation of down-type quarks in the SM mass eigenbasis, and $\lambda_{ij}$ are the flavor-violating couplings between generations. For simplicity, we restrict our analysis to off-diagonal couplings by assuming $\lambda_{ii} = 0$ (i.e., no diagonal interactions). This approach allows us to isolate genuinely flavor-changing effects, which have received comparatively less attention in the literature. Diagonal flavor-conserving interactions have been thoroughly studied and face more stringent constraints, thus we exclude them from our consideration.

To connect the above low-energy effective interaction to a UV completion, we start by considering a simple extension of the SM, where the interaction of $\phi $ with quarks arises from higher-dimensional operators. Before electroweak symmetry breaking (EWSB), the relevant Lagrangian is:
\begin{equation}
-\mathcal{L} \supset y_{u, i j} \bar{Q}_{L, i}^{u, 0} \tilde{H} u_{R, j}^0+\left(y_{d, i j}+\frac{\lambda_{i j}^{\prime} \phi}{\Lambda^{\prime}}\right) \bar{Q}_{L, i}^{d, 0} H d_{R, j}^0+\text { h.c. }
\label{eq:UV}
\end{equation}
where $y_{u/d}$ are the SM Yukawa coupling matrices, $Q^{u/d,0}_L$  are the left-handed $\mathrm{SU}(2)_L$ quark doublets, $u_R^0$ and $d_R^0$ are the right-handed up and down quark singlets, and $H$ is the Higgs doublet. The term $\frac{\lambda_{ij}' \phi}{\Lambda'}$ parameterizes a higher-dimensional operator, which can arise from integrating out some heavier fermions; here $\lambda'_{ij}$  are the coupling constants of the new interaction, and $\Lambda'$ is the scale of new physics responsible for generating this interaction. 

After EWSB, the down-type Yukawa sector induces a $\phi$-dependent contribution to the mass matrix. In the SM down-quark mass eigenstate basis, this can be written as
\begin{equation}\label{eq:yuka}
\begin{aligned}
-\mathcal{L}&\supset \left(1+\frac{h}{v}\right)(\bar{d}_L, \bar{s}_L,\bar{b}_L)\left[\operatorname{diag}\left(m_d, m_s, m_b\right)+  \frac{v \phi}{\sqrt{2}\Lambda'} {V_d^L}^\dagger \left(\begin{array}{ccc}
\lambda_{11}' & \lambda_{12}' &  \lambda_{13}'\\
\lambda_{21}' & \lambda_{22}' &  \lambda_{23}' \\
\lambda_{31}' & \lambda_{32}' & \lambda_{33}'
\end{array}\right)V_d^R\right]
\left(\begin{array}{c}
d_R\\
s_R\\
b_R
\end{array}\right)+\text { h.c. }
\end{aligned}
\end{equation}
where $V_d^{L,R}$ are the unitary matrices that rotate to the down-quark mass basis. The coupling term $\frac{v}{\sqrt{2}\Lambda'} V_d^{L\dagger} \lambda' V_d^R$ is generically non-diagonal. To isolate the novel phenomenology of flavor-violating ULDM, we assume this mass-basis coupling is symmetric with vanishing diagonal entries. Defining the off-diagonal components as $\lambda_{ij}$, the effective Lagrangian becomes:
\begin{equation}
\begin{aligned}
-\mathcal{L} &\supset\left(1+\frac{h}{v}\right)\left(\bar{d}_L, \bar{s}_L, \bar{b}_L\right)\left[\operatorname{diag}\left(m_d, m_s, m_b\right)+\phi\left(\begin{array}{ccc}
0 & \lambda_{12} & \lambda_{13} \\
\lambda_{12} & 0 & \lambda_{23} \\
\lambda_{13} & \lambda_{23} & 0
\end{array}\right)\right]\left(\begin{array}{c}
d_R \\
s_R \\
b_R
\end{array}\right)+ \text{h.c.},\\
&\equiv \left(1+\frac{h}{v}\right)\bar{\bm d}_L M \bm d_R+\text{h.c.}\, .
\end{aligned}
\end{equation}
This is the effective Lagrangian in the down-quark mass basis. Here $\bm d=(d,s,b)^T$ denotes the down-type quark fields. Consequently, for the remainder of this paper, unless otherwise specified, we work with the effective interaction in Eq.~\eqref{eq:eff_lag} and do not return to Eq.~\eqref{eq:UV}.

\subsection{Classical Interpretation}
In the classical (coherent-field or wave) interpretation, the ULDM field in a volume $V$ can be written as
\begin{equation}
    \phi(x)=\sum_{\bf {p}}\sqrt{\frac{2N_{\bf p}}{V \omega_{\bf p}}}\cos(\omega_{\bf p}t-{\bf p}\cdot x+\phi_{\bf p})\, ,
\end{equation}
where the phases $\phi_{\mathbf{p}}$ are random variables uniformly distributed in $[0, 2\pi]$. The occupation number for each mode $\mathbf{p}$ is determined by the local dark matter energy density $\rho_{\rm DM}$ and the velocity distribution $f(\mathbf{p})$ as
$N_{\bf p}=\rho_{\rm DM}V f({\bf p})(\Delta p)^3/\omega_{\bf p}$.
where $f(\mathbf{p})$ follows the standard Maxwell-Boltzmann distribution of the Standard Halo Model \cite{Bovy:2009dr}, normalized such that $\int d^3\mathbf{p} \, f(\mathbf{p}) = 1$.
Locally, within a coherence time, the field can be approximated as spatially homogeneous, with a time dependence given by
\begin{equation}
\phi(t) =\bar{\phi}\cos(m_\phi t+\phi_0)= \frac{\sqrt{2\rho_{\rm DM}}}{m_\phi} \cos(m_\phi t+\phi_0),
\label{DM-eq}
\end{equation}
where $\phi_0$ is a random initial phase and we take the local DM energy density to be $\rho_{\rm DM}=0.4~{\rm GeV\,cm^{-3}}$ \cite{deSalas:2020hbh}.

In this background, the interaction between $\phi$ and down-type quarks induces time-dependent corrections to the quark mass matrix $M$. Including both the SM masses and the ULDM-induced off-diagonal terms, the mass matrix can be diagonalized order by order. In the perturbative regime $|\lambda_{ij}\phi|\ll |m_i-m_j|$ (equivalently $|\lambda_{ij}\phi/(m_i-m_j)|\ll 1$, shown in Fig.~\ref{fig:Combined-constraint}), the mass matrix can be diagonalized by an orthogonal transformation to first order in $\lambda_{ij}\phi$, leading to
\begin{equation}
U \simeq\left(\begin{array}{ccc}
1 & \frac{\lambda_{12} \phi}{m_s-m_d} & \frac{\lambda_{13} \phi}{m_b-m_d} \\
-\frac{\lambda_{12} \phi}{m_s-m_d} & 1 & \frac{\lambda_{23} \phi}{m_b-m_s} \\
-\frac{\lambda_{13} \phi}{m_b-m_d} & -\frac{\lambda_{23} \phi}{m_b-m_s} & 1
\end{array}\right), \quad
U^T M U=\operatorname{diag}\!\left(m_d, m_s, m_b\right)+\mathcal O\!\big((\lambda\phi)^2\big).
\end{equation}
This leads to the mass eigenstates
\begin{equation}
\left(\begin{array}{l}
d^{\prime} \\
s^{\prime} \\
b^{\prime}
\end{array}\right)=U^T\left(\begin{array}{l}
d \\
s \\
b
\end{array}\right) .
\end{equation}
At second order in $\phi$, the down-type quark masses receive corrections given by
\begin{equation}
\begin{aligned}
m_d^{\prime} & \simeq m_d+\phi^2\left(\frac{\lambda_{12}^2}{m_d-m_s}+\frac{\lambda_{13}^2}{m_d-m_b}\right) ,\\
m_s^{\prime} & \simeq m_s+\phi^2\left(\frac{\lambda_{12}^2}{m_s-m_d}+\frac{\lambda_{23}^2}{m_s-m_b}\right) ,\\
m_b^{\prime} & \simeq m_b+\phi^2\left(\frac{\lambda_{13}^2}{m_b-m_d}+\frac{\lambda_{23}^2}{m_b-m_s}\right).
\end{aligned}
\label{eq:quark-mass-change}
\end{equation}
These shifts are of order $\lambda_{ij}^2\phi^2$, arising from a double insertion of the off-diagonal interaction $\lambda_{ij}\phi\,\bar d_i d_j$.
They can be encoded in an effective interaction,
\begin{equation}
\begin{aligned}
\delta \mathcal{L}
&\supset- \sum_{i,j=1,j\neq i}^3\frac{\lambda_{ij}^2}{(m_i-m_j)}  \phi^2 \bar{d}'_i d'_i\\
&= -\sum_{i =1}^3 \frac{1}{\Lambda_i^2} \, m_i\phi^2 \bar{d}'_i d'_i.
\end{aligned}
\end{equation}
Here $\Lambda_i$ labels the effective ULDM coupling to the $i$-th down-type quark and satisfies
$\Lambda_i^{-2}=\sum_{j=1,j\neq i}^3\frac{\lambda_{ij}^2}{m_i(m_i-m_j)}$.
With this, the effective scalar--nucleon interaction can be expressed as
\begin{equation}\label{eq:eff_p_n}
\begin{aligned}
    \mathcal{L}_{\rm eff}\supset -\frac{1}{\Lambda_p}\phi^2\bar{p}p- \frac{1}{\Lambda_n}\phi^2 \bar{n}n,
\end{aligned}
\end{equation}
where the coupling strengths are derived following Refs.~\cite{Cheng:2012qr,Bishara:2017pfq} as
\begin{equation}\label{eq:Lambda_p_n}
\begin{aligned}
    \frac{1}{\Lambda_p}&=\frac{\sigma_d^p}{\Lambda_d^2} + \frac{\sigma_s^p}{\Lambda_s^2} + \frac{2}{27}\frac{m_p-\sigma_u^p-\sigma_d^p-\sigma_s^p}{\Lambda_b^2},\\
   \frac{1}{\Lambda_n}&= \frac{\sigma_d^n}{\Lambda_d^2} + \frac{\sigma_s^n}{\Lambda_s^2} + \frac{2}{27}\frac{m_n-\sigma_u^n-\sigma_d^n-\sigma_s^n}{\Lambda_b^2},
\end{aligned}
\end{equation}
with $\sigma_u^p=17\pm 5~\text{MeV}$, $\sigma_d^p=32\pm 10~\text{MeV}$, $\sigma_u^n=15\pm 5~\text{MeV}$, $\sigma_d^n=36\pm 10~\text{MeV}$, and $\sigma_s^{p/n}=41.3\pm7.7~\text{MeV}$ \cite{Junnarkar:2013ac,Yang:2015uis,Durr:2015dna,RuizdeElvira:2017stg,Hoferichter:2015dsa,Alvarez-Ruso:2014sma,Abdel-Rehim:2016won,Hoferichter:2016ocj,Crivellin:2013ipa}.

In the SM quark mass basis, the quark kinetic terms and electroweak interactions read
\begin{equation}
\begin{aligned}
&\mathcal{L}\supset~ \bar{\bm u}_L i \gamma^\mu \partial_\mu  \bm u_L+\bar{\bm d}_L i \gamma^\mu \partial_\mu \bm d_L+\bar{\bm u}_R i \gamma^\mu \partial_\mu \bm u_R + \bar{\bm d}_R i \gamma^\mu \partial_\mu \bm d_R\\
&~-\frac{g}{\sqrt{2}} \bar{\bm u}_L V_{\rm CKM} \gamma^\mu W_\mu^{+} \bm d_L+\text{h.c.}\\
&~-\frac{g}{c_w}Z_\mu\left(\left(\frac{1}{2}-\frac{2}{3} \sin ^2 \theta_w\right) \bar{\bm u}_L \gamma^\mu \bm u_L-\frac{2}{3} \sin ^2 \theta_w \bar{\bm u}_R \gamma^\mu \bm u_R + \left(-\frac{1}{2}+\frac{1}{3} \sin ^2 \theta_w\right) \bar{\bm d}_L \gamma^\mu \bm d_L+\frac{1}{3} \sin ^2 \theta_w \bar{\bm d}_R \gamma^\mu \bm d_R\right)\\
&~-\frac{2}{3}e A_\mu\bar{\bm u} \gamma^\mu \bm u + \frac{1}{3}e A_\mu\bar{\bm d} \gamma^\mu \bm d,
\end{aligned}
\end{equation}
where $V_{\rm CKM}$ is the CKM matrix \cite{Cabibbo:1963yz,Kobayashi:1973fv}.
After the $\phi$-induced mixing, the kinetic term and the charged-current interaction receive corrections. The corrected Lagrangian can be written as
\begin{equation}\label{eq:kine}
\begin{aligned}
\mathcal{L}&\supset \bar{\bm d}'\, i \gamma^\mu \left(U^T \partial_\mu U+\partial_\mu\right) \bm d'
+ \left[- \frac{g}{\sqrt{2}} \bar{\bm u}_L^{\prime} V_{\rm CKM} U \gamma^\mu W_\mu^{+} \bm d_L'+\text{h.c.}\right],
\end{aligned}
\end{equation}
where the up-type quarks remain unchanged, $u_L' = u_L$.
The kinetic correction induces a derivative interaction among different down-type quarks. To leading order in $\lambda_{ij}\phi$, one finds
\begin{equation}\label{eq:kin}
    \mathcal{L}_{\rm kin}\supset i g_{ij}(\partial_\mu\phi) \bar{d}'_i\gamma^\mu d'_j,
\end{equation}
with
\begin{equation}
g_{ij} = \begin{pmatrix}
0& -\frac{\lambda_{12}}{m_s-m_d} &-\frac{\lambda_{13}}{m_b-m_d}\\
\frac{\lambda_{12}}{m_s-m_d}& 0 & -\frac{\lambda_{23}}{m_b-m_s}\\
\frac{\lambda_{13}}{m_b-m_d} & \frac{\lambda_{23}}{m_b-m_s}&0
\end{pmatrix}.
\end{equation}
Since $\phi$ is a ULDM background field, $\partial_\mu\phi$ is characterized by a typical four-momentum
$k_\mu\sim(m_\phi,\,m_\phi \vec v_\phi)$ with $|\vec v_\phi|\sim 10^{-3}$.
The derivative interaction in Eq.~\eqref{eq:kin} is therefore additionally suppressed by $m_\phi$ and is subdominant compared with the mass-mixing and CKM effects; we neglect it following Ref.~\cite{Bigaran:2025uzn}.

For the charged-current interaction, we define the modified CKM matrix as $V_{\mathrm{CKM}}^{\prime}=V_{\mathrm{CKM}} U$, which can be expressed explicitly as
\begin{align}\label{eq:CKM-matrix} 
V_{\rm CKM}' \simeq 
\begin{pmatrix}
V_{ud} - \frac{V_{us} \phi \lambda_{12}}{-m_d + m_s} - \frac{V_{ub} \phi \lambda_{13}}{m_b - m_d} &
V_{us} - \frac{V_{ud} \phi \lambda_{12}}{m_d - m_s} - \frac{V_{ub} \phi \lambda_{23}}{m_b - m_s} &
V_{ub} - \frac{V_{ud} \phi \lambda_{13}}{-m_b + m_d} - \frac{V_{us} \phi \lambda_{23}}{-m_b + m_s} \\
V_{cd} - \frac{V_{cs} \phi \lambda_{12}}{-m_d + m_s} - \frac{V_{cb} \phi \lambda_{13}}{m_b - m_d} &
V_{cs} - \frac{V_{cd} \phi \lambda_{12}}{m_d - m_s} - \frac{V_{cb} \phi \lambda_{23}}{m_b - m_s} &
V_{cb} - \frac{V_{cd} \phi \lambda_{13}}{-m_b + m_d} - \frac{V_{cs} \phi \lambda_{23}}{-m_b + m_s} \\
V_{td} - \frac{V_{ts} \phi \lambda_{12}}{-m_d + m_s} - \frac{V_{tb} \phi \lambda_{13}}{m_b - m_d} &
V_{ts} - \frac{V_{td} \phi \lambda_{12}}{m_d - m_s} - \frac{V_{tb} \phi \lambda_{23}}{m_b - m_s} &
V_{tb} - \frac{V_{td} \phi \lambda_{13}}{-m_b + m_d} - \frac{V_{ts} \phi \lambda_{23}}{-m_b + m_s}.
\end{pmatrix}
\end{align}
All CKM elements therefore receive $\mathcal{O}(\lambda_{ij}\phi)$ corrections in the coherent-field limit, potentially inducing observable flavor-violating effects.

\begin{figure}[htbp]
    \centering
    \includegraphics[width=0.95\linewidth]{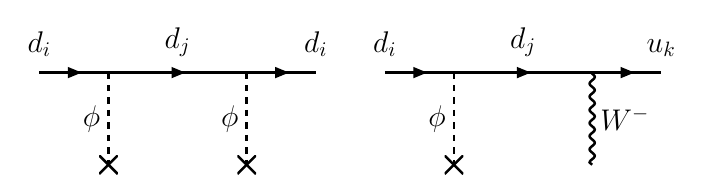}
    \captionsetup{justification=raggedright}
    \caption{Two equivalent viewpoints: re-diagonalization in the presence of a ULDM background (mass mixing), and perturbation theory in the SM mass basis.}
    \label{fig:feynman}
\end{figure}

These results can be corroborated by a direct calculation in the SM mass basis. To establish this equivalence explicitly, we consider the self-energy and vertex corrections shown in Fig.~\ref{fig:feynman}. In the non-relativistic limit where $m_\phi \ll m_i$ and the field momentum is negligible ($m_\phi |\vec{v}_\phi| \ll m_i$), the amplitude in the left panel reduces via the Dirac equation to:
\begin{equation}
\begin{aligned}  \label{eq:massequivalent}
    -i\lambda_{ij}^2 \phi^2 \bar{d}_i (p) \frac{1}{\slashed{p}-m_j} d_i(p) &= -i \lambda_{ij}^2 \phi^2 \bar{d}_i (p) \frac{\slashed{p}+m_j}{m_i^2-m_j^2} d_i(p)\\
    &\simeq -i \phi^2 \frac{\lambda_{ij}^2 }{m_i-m_j}  \bar{d}_i(p)  d_i(p).
\end{aligned}
\end{equation}
Similarly, for the charged-current amplitude shown in the right panel, one finds
\begin{equation}
\begin{aligned}\label{eq:CKMChangeEquivalent}
    -i\lambda_{ij} V_{\text{CKM},kj} \bar{u}_{L,k}(p_k) \gamma^\mu  \epsilon_\mu^-(p_W)  \frac{1}{\slashed{p}_i-m_j} d_{L,i} (p_i) & \simeq -i\lambda_{ij} V_{\text{CKM},kj} \bar{u}_{L,k}(p_k) \gamma^\mu  \epsilon_\mu^-(p_W)  \frac{\slashed{p}+m_j}{m_i^2-m_j^2} d_{L,i} (p_i)\\
    & = \frac{-i\lambda_{ij}}{m_i-m_j}\cdot V_{\text{CKM},kj} \bar{u}_{L,k}(p_k) \gamma^\mu  \epsilon_\mu^-(p_W)  d_{L,i} (p_i).
\end{aligned}
\end{equation}
The agreement between Eq.~\eqref{eq:quark-mass-change} (Eq. \eqref{eq:CKM-matrix}) and Eq.~\eqref{eq:massequivalent} (Eq. \eqref{eq:CKMChangeEquivalent}) confirms that the perturbative rotation and the diagrammatic expansion are consistent at leading order.
For the derivative term in Eq.~\eqref{eq:kin}, integrating by parts and using the Dirac equation gives
\begin{equation}
\begin{aligned}
    \mathcal{L}_{\rm kin}&\supset i g_{ij}(\partial_\mu\phi) \bar{d}'_i\gamma^\mu d'_j
     = -i g_{ij} \phi \partial_\mu (\bar{d}'_i\gamma^\mu d'_j)\\
    & = - g_{ij} \phi (m_j' - m_i') \bar{d}'_i d'_j
     \simeq -\lambda_{ij} \phi \bar{d}_i d_j,
\end{aligned}
\end{equation}
where the last step keeps only the leading-order contribution. This clarifies the physical meaning of the derivative term as a rewriting of the original Yukawa interaction at leading order.

In summary, we have established a quark flavor-violating ULDM setup in the coherent-field limit. The dominant effects are the induced mass mixing in the down sector and the resulting time-dependent corrections to the CKM matrix, which can be probed by current and upcoming experiments.

\subsection{Quantum Interpretation}
In the previous subsection, the ULDM field $\phi$ was treated as a coherent classical background. In this subsection, we present the quantum-field interpretation, in which $\phi$ is treated as a dynamical quantum field, and its effects are computed perturbatively in the SM mass eigenbasis.

In this framework, no field redefinition or re-diagonalization of the quark mass matrix is required. Instead, all physical effects arise from insertions of the flavor-violating interaction
\begin{equation}
    \mathcal{L}_{\rm int}=\phi\sum_{i,j=1,i\neq j}^3\lambda_{ij}\bar{d}_i d_j,
\end{equation}
written directly in the SM mass eigenstates. At the level of a single insertion, this operator mediates flavor-changing transitions between down-type quarks. At the parton level, the corresponding decay width is
\begin{equation}
\begin{aligned}
    \Gamma(d_i\to d_j \phi)&=\frac{N_c \lambda_{ij}^2 (m_i-m_j)(m_i+m_j)^3}{16\pi m_i^3}
    \simeq\frac{N_c \lambda_{ij}^2 m_i}{16\pi},
\end{aligned}
\end{equation}
where $N_c=3$ is the color factor and the approximation holds for $m_i \gg m_j$, neglecting the ULDM mass $m_\phi$.

\begin{figure}[htbp]
\centering
    \includegraphics[width=0.98\linewidth]{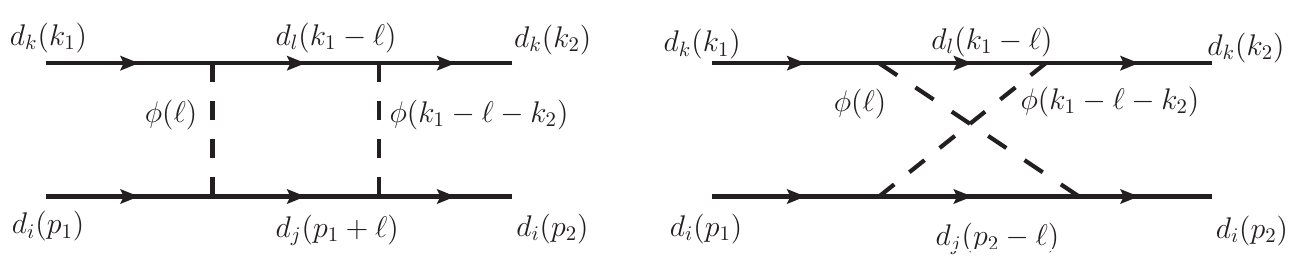}
    \captionsetup{justification=raggedright}
    \caption{Representative one-loop diagram for the induced fifth force: $d_i(p_1)d_k(k_1)\to d_i(p_2)d_k(k_2)$ via exchange of two virtual mediators $\phi$, with $d_i$ the $i$th generation of down-type quarks and $p_i$, $k_i$, $\ell$ the momenta.}
    \label{fig:1-loop-feynman}
\end{figure}

For the exotic fifth force, the absence of diagonal couplings implies that the leading contribution arises at one loop. A representative diagram is shown in Fig.~\ref{fig:1-loop-feynman}. To compute the induced fifth force between fermions $d_i$ and $d_k$, one may evaluate the one-loop scattering amplitude with two off-shell scalars $\phi$ and two virtual fermions $d_j$ and $d_{\ell}$,
\begin{equation}
\begin{aligned}
    i\mathcal{M}=&\int \frac{d^4\ell}{(2\pi)^4}\frac{1}{(\ell^2-m_\phi^2)\big((k_1-\ell-k_2)^2-m_\phi^2\big)}\\
    &\cdot\left[\bar{u}_{d_i}(p_2,m_i)\,\lambda_{ij}\frac{i}{\slashed{p}_1+\slashed{\ell}-m_j}\lambda_{ji}\,u_{d_i}(p_1,m_i)\times \bar{u}_{d_k}(k_2,m_k)\,\lambda_{k\ell}\frac{i}{\slashed{k}_1-\slashed{\ell}-m_\ell}\lambda_{\ell k}\,u_{d_k}(k_1,m_k)\right.\\
    &~~~~~+\left.\bar{u}_{d_i}(p_2,m_i)\,\lambda_{ij}\frac{i}{\slashed{p}_2-\slashed{\ell}-m_j}\lambda_{ji}\,u_{d_i}(p_1,m_i)\times \bar{u}_{d_k}(k_2,m_k)\,\lambda_{k\ell}\frac{i}{\slashed{k}_1-\slashed{\ell}-m_\ell}\lambda_{\ell k}\,u_{d_k}(k_1,m_k)\right].
\end{aligned}
\end{equation}
In the non-relativistic limit relevant for fifth-force experiments, the external fermions are nearly on shell. The long-range potential is dominated by loop momenta $\ell\sim m_\phi$, while contributions from $\ell\gg m_\phi$ reduce to contact interactions. Using the Dirac equation and expanding in the small momentum transfer, one finds
\begin{equation}
\begin{aligned}
    i\mathcal{M}&\simeq \int \frac{d^4\ell}{(2\pi)^4}\frac{1}{(\ell^2-m_\phi^2)\big((k_1-\ell-k_2)^2-m_\phi^2\big)}\\
    &\cdot\left[\frac{\lambda_{ij}\lambda_{ji}}{m_i-m_j}\bar{u}_{d_i}(p_2,m_i)\, u_{d_i}(p_1,m_i) \times \frac{\lambda_{k\ell}\lambda_{\ell k}}{m_k-m_\ell}\bar{u}_{d_k}(k_2,m_k)\,u_{d_k}(k_1,m_k)\right.\\
    &~~~+\left.\frac{\lambda_{ij}\lambda_{ji}}{m_i-m_j}\bar{u}_{d_i}(p_2,m_i)\,u_{d_i}(p_1,m_i)\times \frac{\lambda_{k\ell}\lambda_{\ell k}}{m_k-m_\ell}\bar{u}_{d_k}(k_2,m_k)\,u_{d_k}(k_1,m_k)\right]
    +\mathcal{O}\left(\frac{\ell}{(m_{i/k}-m_{j/\ell})^2}\right)\\
    =&\int \frac{d^4\ell}{(2\pi)^4}\frac{2}{(\ell^2-m_\phi^2)\big((k_1-\ell-k_2)^2-m_\phi^2\big)}\\
    &\times\left[\frac{\lambda_{ij}\lambda_{ji}}{m_i-m_j}\bar{u}_{d_i}(p_2,m_i)\, u_{d_i}(p_1,m_i)\cdot \frac{\lambda_{k\ell}\lambda_{\ell k}}{m_k-m_\ell}\bar{u}_{d_k}(k_2,m_k)\,u_{d_k}(k_1,m_k)\right]
    +\mathcal{O}\left(\frac{\ell}{(m_{i/k}-m_{j/\ell})^2}\right),
\end{aligned}
\end{equation}
where the limit $|\ell|\ll m_i$ has been applied.

The derivation is straightforward. For example, for the $d_i$ line,
\begin{equation}
\begin{aligned}
    &\bar{u}_{d_i}(p_2,m_i)\lambda_{ij}\frac{i}{\slashed{p}_1+\slashed{\ell}-m_j}\lambda_{ji}u_{d_i}(p_1,m_i)
    =i\lambda_{ij}\lambda_{ji}\,
    \bar{u}_{d_i}(p_2,m_i)\frac{\slashed{p}_1+\slashed{\ell}+m_j}{(p_1+\ell)^2-m_j^2}\,u_{d_i}(p_1,m_i)\\
    &\simeq \frac{i\lambda_{ij}\lambda_{ji}}{m_i-m_j}\,\bar{u}_{d_i}(p_2,m_i)\,u_{d_i}(p_1,m_i)
    +\mathcal{O}\!\left(\frac{\ell}{(m_i-m_j)^2}\right).
\end{aligned}
\end{equation}
It is therefore convenient to integrate out the virtual fermions and combine the two Yukawa vertices into a higher-dimensional effective operator. At leading order, the resulting interaction is
\begin{equation}
\begin{aligned}
    \mathcal{L}_{\rm eff}&\simeq -\sum_{i,j=1,j\neq i}^3\frac{\lambda_{ij}\lambda_{ji}}{(m_i-m_j)}\,\bar{d}_i d_i \,\phi^2
    =-\sum_{i=1}^3 \frac{1}{\Lambda_i^2}\, m_i \phi^2 \bar{d}_i d_i,
\end{aligned}
\end{equation}
This interaction has the same structure as the quark mass corrections in Eq.~\eqref{eq:quark-mass-change} and provides a convenient starting point for computing the induced fifth force between nuclei. As an intermediate step, we match onto scalar--nucleon interactions,
\begin{equation}\label{eq:eff_p_n2}
\begin{aligned}
    \mathcal{L}_{\rm eff}\supset- \frac{1}{ \Lambda_p}\phi^2\bar{p}p- \frac{1}{ \Lambda_n}\phi^2 \bar{n}n,
\end{aligned}
\end{equation}
where the coupling strengths, following Refs.~\cite{Cheng:2012qr,Bishara:2017pfq}, take the same form as Eq.~\eqref{eq:Lambda_p_n}. With these interactions, the exotic fifth force between two atoms mediated by quadratic $\phi^2$ exchange can be written as \cite{Banks:2020gpu, Bauer:2023czj}
\begin{equation}\label{eq:fifth-force}
F_{12}(r)=\left(N_{1,p}\frac{1}{\Lambda_{p}}+N_{1, n}\frac{1}{\Lambda_{n}}\right)\left(N_{2,p}\frac{1}{\Lambda_{p}}+N_{2,n}\frac{1}{\Lambda_{n}}\right)\frac{m_\phi}{4\pi^3 r^3}\left(\frac{3}{2}K_1(2m_\phi r)+m_\phi r K_0(2m_\phi r)\right),
\end{equation}
where $N_{i,p}$ and $N_{i,n}$ are the numbers of protons and neutrons in the $i$th atom, $r$ is the separation between the two atoms, and $K_n$ denotes the $n$th modified Bessel function of the second kind.
Note that the force scales as $r^{-3}$ rather than $r^{-2}$, since it arises from quadratic exchange ($\phi^2$) rather than single-$\phi$ exchange.
These expressions will be used in Sec.~\ref{sec:particle} to derive experimental constraints on the model.

\section{Constraints on ULDM as a Coherent Background Field}
\label{sec:uldm}

As discussed above, the ULDM field $\phi$ can be treated as a coherently oscillating classical background. Such a background induces flavor mixing among down-type quarks, leading to periodic modulations of the quark masses and of the CKM matrix elements. In this section, we study these two characteristic signatures using direct CKM measurements, nuclear decays, atomic clock comparisons, and other related precision observables.

\subsection{Direct Measurement of CKM Matrix}
As shown in Eq.~\eqref{eq:CKM-matrix}, in the presence of a ULDM background the CKM elements receive an oscillatory correction. In this subsection, we derive an estimate of the sensitivity by requiring that the ULDM-induced modulation does not exceed the current experimental uncertainty. We stress that a dedicated constraint would require the experiment-specific time binning, correlations, and the full global-fit machinery, which is beyond the scope of this subsection.

To obtain a meaningful constraint, the characteristic experimental time resolution $\Delta t_{\exp}^{\rm res}$ is crucial. Only when $\Delta t_{\exp}^{\rm res}$ is shorter than the ULDM oscillation period $\tau_\phi=2\pi/m_\phi$, and the total running time is sufficiently long, can the measurements be sensitive to an oscillatory signal of this type. Using the PDG global-fit CKM magnitudes and their quoted $1\sigma$ uncertainties~\cite{CKMmatrix}, which we denote as $|V_{ij}|=\bar V_{ij}\pm \delta V^{\rm exp}_{ij}$, we take
\begin{align}\label{eq:CKM measured}
\left| V_{\mathrm{CKM}} \right| =
\begin{pmatrix}
0.97435 \pm 0.00016 & 0.22501 \pm 0.00068 & 0.003732^{+0.000090}_{-0.000085} \\
0.22487 \pm 0.00068 & 0.97349 \pm 0.00016 & 0.04183^{+0.00079}_{-0.00069} \\
0.00858^{+0.00019}_{-0.00017} & 0.04111^{+0.00077}_{-0.00068} & 0.999118^{+0.000029}_{-0.000034}
\end{pmatrix}.
\end{align}

In the absence of a time-resolved global fit, we adopt a conservative order-of-magnitude criterion and impose the element-wise requirement
$|\delta V^{\rm ULDM}_{ij}|\lesssim \delta V^{\rm exp}_{ij}$
\footnote{Here $\delta V^{\rm ULDM}_{ij}$ denotes the (model-dependent) amplitude of the ULDM-induced oscillatory correction to $|V_{ij}|$, while $\delta V^{\rm exp}_{ij}$ is the corresponding PDG uncertainty. This criterion is intended only as a conservative estimate and does not account for correlations or experimental time-binning.}.
Solving the resulting linear inequalities for the three independent couplings, we obtain approximate bounds on the combinations $\lambda_{ij}\phi$:
$\lambda_{12}\phi \lesssim 6.22\times10^{-5}\,{\rm GeV}$,
$\lambda_{13}\phi \lesssim 3.63\times10^{-4}~{\rm GeV}$,
and $\lambda_{23}\phi \lesssim 1.54\times10^{-3}\,{\rm GeV}$.

Finally, treating ULDM as a coherently oscillating classical field with amplitude $\bar\phi\simeq \sqrt{2\rho_{\rm DM}}/m_\phi$ (see Eq.~\eqref{DM-eq}), we translate these limits into indicative constraints on $\lambda_{ij}$:
$\lambda_{12}\lesssim \mathcal{O}(10^{-11})\left(\frac{m_\phi}{10^{-18}\,{\rm eV}}\right)$,
$\lambda_{13}\lesssim \mathcal{O}(10^{-10})\left(\frac{m_\phi}{10^{-18}\,{\rm eV}}\right)$,
and $\lambda_{23}\lesssim \mathcal{O}(10^{-9})\left(\frac{m_\phi}{10^{-18}\,{\rm eV}}\right)$.
These bounds should be regarded as indicative only, since they do not incorporate the experiment-dependent time resolution, sampling strategy, or the full structure of the CKM global fit.

\subsection{Nuclear Beta Decay}
ULDM behaves as a coherently oscillating classical field and can induce time-dependent variations in the SM parameters to which it couples, including CKM matrix elements. We focus on the element $V_{ud}$, whose most precise determinations rely on nuclear beta decay. Unlike the previous case, where the time information was left unspecified, time-resolved decay measurements enable a more targeted search for ULDM-induced oscillatory signatures.

Nuclear beta decay is particularly well suited for this purpose due to the long lifetimes of many beta-unstable nuclei and the high precision achievable in decay-rate measurements. Since the decay width depends directly on $V_{ud}$, even small temporal modulations can, in principle, be probed. In contrast, analogous searches involving other CKM elements are strongly limited by the extremely short lifetimes of the relevant processes. The element $V_{ud}$ is extracted from superallowed $0^{+}\to 0^{+}$ nuclear beta decays, for which nuclear and radiative corrections are conventionally absorbed into a single corrected comparative half-life~\cite{Seng:2025hnz}. Consequently, precision half-life measurements provide direct and stringent sensitivity to variations in $V_{ud}$.

In the presence of ULDM, we parametrize the time dependence of $V_{ud}$ as
\begin{equation}
V_{ud}(t)=V_{ud}+\delta V_{ud}\cos(m_\phi t),
\end{equation}
with
\begin{equation}
\delta V_{ud}=-\left(\frac{V_{us}\lambda_{12}}{m_s-m_d}+\frac{V_{ub}\lambda_{13}}{m_b-m_d}\right)\frac{\sqrt{2\rho_{\mathrm{DM}}}}{m_\phi}\, .
\end{equation}
The ULDM-induced modulation of $V_{ud}$ leads to a time-dependent nuclear decay width,
\begin{equation}
\begin{aligned}
\Gamma(t)
&=\Gamma_0\left[1+\frac{2\delta V_{ud}}{V_{ud}}\cos(m_\phi t)+\left(\frac{\delta V_{ud}}{V_{ud}}\right)^2\cos^2(m_\phi t)\right]\\
&\simeq \Gamma_0\left[1+\frac{2\delta V_{ud}}{V_{ud}}\cos(m_\phi t)+\mathcal{O}\!\left(\frac{\delta V_{ud}^2}{V_{ud}^2}\right)\right],
\end{aligned}
\end{equation}
where $\Gamma_0$ denotes the SM decay width, while the additional terms arise from the ULDM-induced correction and correspond to the linear and quadratic contributions in $\delta V_{ud}/V_{ud}$. 
For $|\delta V_{ud}| \ll |V_{ud}|$, the linear term provides the leading modulation. This contribution can be resolved provided that the experimental time resolution is sufficient compared with the ULDM oscillation period. By contrast, if the ULDM period is much shorter than the experimental time resolution, the linear modulation averages to zero, and the quadratic correction becomes the leading residual effect.

In the present analysis, we only retain the modulation induced via $V_{ud}$ as an illustrative example to demonstrate the methodological feasibility of the time-dependent approach. We emphasize that this is not intended as a fully rigorous treatment, given the subtleties involved in the $Q$-value analysis. In particular, ULDM-induced shifts in the proton and neutron masses can also modify the decay $Q$-value of ${}^{37}\mathrm{K}$ and thus the phase-space factor governing the lifetime. This would introduce an additional oscillatory contribution to the decay width, typically at frequency $2m_\phi$ in the coherent-field regime. A complete and consistent treatment would require modeling the response of nuclear binding energies to the ULDM background, which lies beyond the scope of the present work and is deferred to future study.

Solving the decay equation $\mathrm{d}N/\mathrm{d}t=-\Gamma(t)\,N$, the nuclear population evolves as
\begin{equation}
\begin{aligned}
N(t)
&=N_0\,\exp(-\Gamma_0 t)\exp\!\left[-\frac{\Gamma_0}{m_\phi}\frac{2\delta V_{ud}}{V_{ud}}\sin(m_\phi t)\right]\\
&\simeq N_0\,\exp(-\Gamma_0 t)\left[1-\frac{\Gamma_0}{m_\phi}\frac{2\delta V_{ud}}{V_{ud}}\sin(m_\phi t)\right],
\end{aligned}
\end{equation}
where $N_0$ is the initial number of nuclei at the start of the experiment, and the second line holds for $\frac{\Gamma_0}{m_\phi}\frac{2\delta V_{ud}}{V_{ud}}\ll 1$. Consequently, the particle counts exhibit a small deviation from the standard exponential decay law, which can be quantified as
\begin{equation}\label{eq:deltaN_over_N}
\frac{\delta N(t)}{N_0\exp(-\Gamma_0 t)}\equiv \frac{\delta N(t)}{N(t)}
\simeq -\frac{\Gamma_0}{m_\phi}\frac{2\delta V_{ud}}{V_{ud}}\sin(m_\phi t)
= A_N \sin(m_\phi t).
\end{equation}
The oscillation amplitude $A_N$ directly encodes the ULDM-induced modulation of $V_{ud}$. An experimental upper limit $A_N^{\max}$ therefore implies
\begin{equation}
\delta V_{ud}<\frac{m_\phi}{2\Gamma_0}\,A_N^{\max}V_{ud},
\end{equation}
which can be translated into bounds on the couplings $\lambda_{ij}$ via Eq.~\eqref{eq:CKM-matrix}.

Existing high-statistics nuclear decay measurements can be directly applied to search for ULDM-induced
oscillations. In particular, the beta decay of ${ }^{37}\mathrm{K}$ studied in Ref.~\cite{Shidling:2014ura},
with sampling interval $\Delta t=50~\mathrm{ms}$ ($f_s=20~\mathrm{Hz}$) and total duration $T\simeq
10~\mathrm{s}$, provides a nominal frequency resolution $\Delta f\simeq 1/T\simeq 0.1~\mathrm{Hz}$ and a
Nyquist limit $f_{\rm Nyq}=f_s/2\simeq 10~\mathrm{Hz}$. To suppress low-frequency systematics , we apply a fourth-order Butterworth high-pass filter with cutoff
$f_c=1~\mathrm{Hz}$ to the measured series $\delta(t)=\delta N(t)/\sqrt{N(t)}$ using zero-phase
filtering. Consequently the analysis focuses on the $1$--$10~\mathrm{Hz}$ band (resolution $\sim
0.1~\mathrm{Hz}$), corresponding to $m_\phi\sim 10^{-15}$--$10^{-14}~\mathrm{eV}$. No statistically
significant oscillatory signal is observed, enabling constraints on ULDM couplings.
The noise standard deviation $\sigma_{\mathrm{filtered}}$ is computed from the filtered series
$\delta(t)=\delta N(t)/\sqrt{N(t)}$. The data are then converted to the
analysis domain via $\delta N(t)/N(t)=\delta_{\rm filtered}(t)/\sqrt{N(t)}$
for further analysis.

\begin{figure}[htb]
    \centering
    \includegraphics[width=0.5\linewidth]{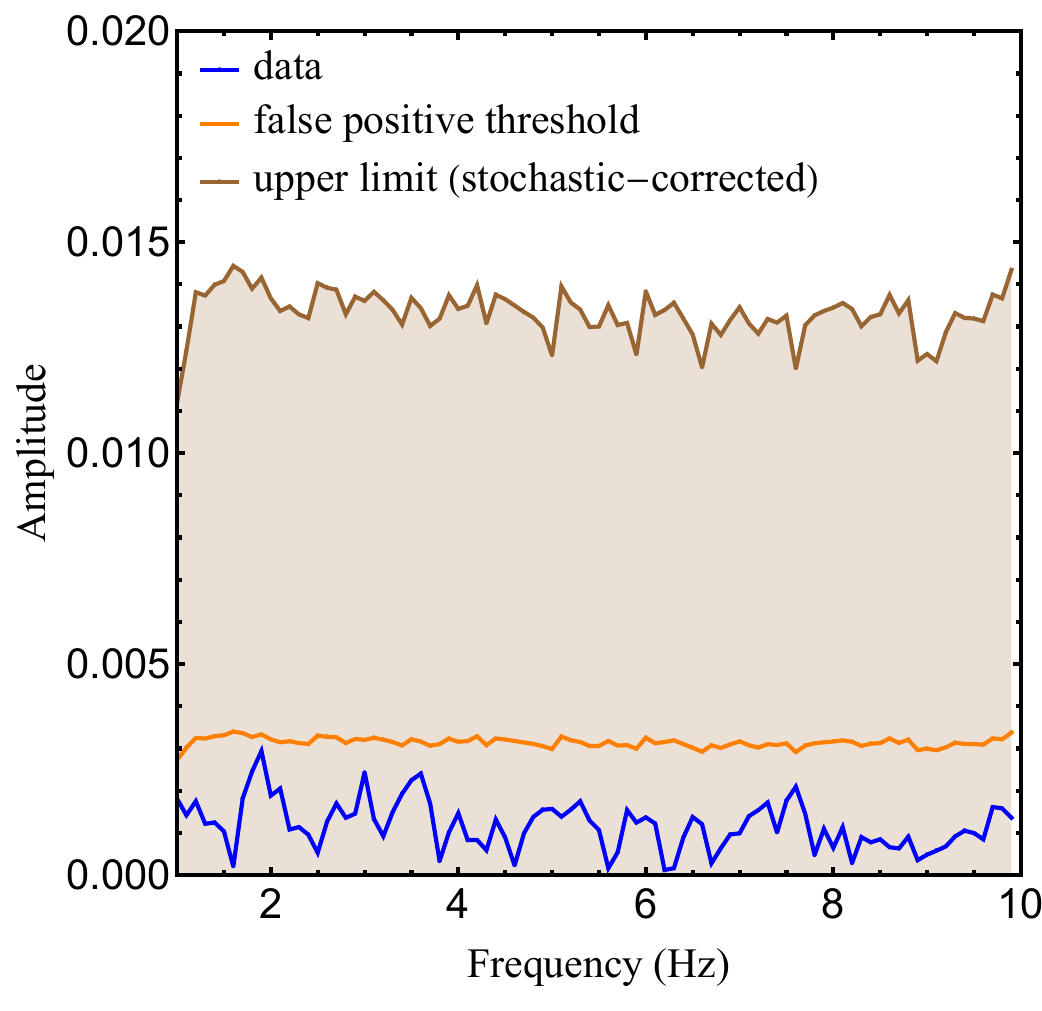}
    \captionsetup{justification=raggedright}
    \caption{Periodogram corresponding to the dataset, along with the MC-derived $95\%$ confidence level threshold and upper limit. The data are compatible with the null hypothesis.}
    \label{fig:MC95}
\end{figure}

Under the null hypothesis that no ULDM signal is present, the residual time-series data points follow a Gaussian distribution around $\delta N/\sqrt{N}=0$. For the background-only case, we generate $10^4$ synthetic time series with the same temporal spacing in the $\delta N/\sqrt{N}$ domain by drawing each point from a zero-mean Gaussian with width $\sigma_{\rm filtered}$, apply the same high-pass filter, and then convert to the analysis domain via $\delta N(t)/N(t)=\delta_{\rm filtered}(t)/\sqrt{N(t)}$. The ensemble of periodograms is used to construct the cumulative distribution function (CDF) of the spectral power at each frequency, from which the $95\%$ confidence-level (CL) false-positive threshold is determined. To account for the look-elsewhere effect, this threshold is corrected using the global trials factor,
\begin{equation}
p_{\text{global}}=1-\left(1-p_{\text{local}}\right)^{N_f},
\end{equation}
where $p_{\text{local}}$ is the local $p$-value at a single trial frequency and $N_f=90$ is the number of
independent trial frequencies after applying the $1~\mathrm{Hz}$ high-pass
cut~\cite{Abel:2022vfg,Wu:2019exd}. As shown in Fig.~\ref{fig:MC95}, the experimental dataset does not exceed
the $95\%$ CL false-positive threshold at any frequency, indicating consistency with the null hypothesis.

To determine exclusion limits on the ULDM couplings, we extend the MC procedure by injecting a ULDM-induced oscillatory signal of fixed amplitude and frequency into each synthetic dataset. For each trial frequency, $5\times10^4$ synthetic noise series are generated in the $\delta N/\sqrt{N}$ domain with the same temporal spacing as the data, drawing each point from a zero-mean Gaussian of width $\sigma_{\rm filtered}$. 
We subsequently inject a sinusoidal signal with a random phase, defined in the $\delta N/N$ domain, and transform the combined series to the analysis domain via $\delta N(t)/N(t)=\delta(t)/\sqrt{N(t)}$ for fast Fourier transform analysis.
The noise plus signal realization is then passed through the same fourth-order Butterworth high-pass filter with cutoff $f_c=1~\mathrm{Hz}$ used for the data.
The $95\%$ CL upper limit on the signal amplitude at each frequency is then determined following the procedure of Refs.~\cite{Centers:2019dyn,Zhang:2023lem}. In the regime $t_{\exp}\ll\tau_{\rm coh}$, a stochastic-to-deterministic correction is applied as in Refs.~\cite{Centers:2019dyn,Zhang:2023lem}. The derived exclusion limits on $\lambda_{12}$ and $\lambda_{13}$ as a function of $m_\phi$ are summarized in Fig.~\ref{fig:Combined-constraint}. Near $m_\phi=4\times10^{-14}~\mathrm{eV}$, the analysis sets $95\%$ CL upper limits $\lambda_{12}\lesssim 10^{-5}$ and $\lambda_{13}\lesssim 10^{-3}$.

\subsection{Tritium Decays}
In addition to the ${}^{37}\mathrm{K}$ decay case, long-term measurements of tritium decay provide another stringent constraint on our model over a complementary range of frequencies. In particular, null results from a 12-year tritium decay dataset have already placed strong limits on ULDM couplings~\cite{Zhang:2023lem,tritiumdecay}. As shown in Refs.~\cite{Dai:2025von,Zhang:2023lem}, ULDM induces coherent oscillations in the down-quark Yukawa sector, leading to time-dependent shifts in the neutron--proton mass difference, nuclear binding energies, and, consequently, the tritium decay rate. These shifts oscillate in synchrony with the ULDM field, providing a distinctive signature of ULDM interactions in nuclear processes.

According to the calculation in Ref.~\cite{Faessler:2011zz}, the total $\beta$-decay rate is
\begin{equation}
\Gamma=\frac{1}{2\pi^3}m_e(G_\beta m_e^2)^2\left(B_\mathrm{F}(^3H)+B_{\mathrm{GT}}(^3H)\right)I^\beta(^3H),
\end{equation}
where $G_\beta\propto V_{ud}$ and $I^\beta$ is the phase-space integral. Following the discussion in Ref.~\cite{Zhang:2023lem}, the fractional change in the decay rate can be written as
\begin{equation}\label{eq:betaGammChange}
I_0\equiv\frac{\delta\Gamma}{\Gamma}
=\frac{\delta I^\beta}{I^\beta}+2\frac{\delta V_{ud}}{V_{ud}}.
\end{equation}

The effect of ULDM on standard nuclear interactions can be described using the pion--nucleon chiral Lagrangian at next-to-leading order~\cite{Bernard:1996gq}. Specifically, the presence of ULDM modifies the proton--neutron mass difference (cf.~Eq.~\eqref{eq:eff_p_n}), which we write as
\begin{equation}
(m_n-m_p)(\phi)=1.57~\mathrm{MeV}
+\Bigg[-7.19\times10^{-9}(\lambda_{12}\phi)^2-1.53\times10^{-10}(\lambda_{13}\phi)^2+2.47\times10^{-15}(\lambda_{23}\phi)^2\Bigg]\mathrm{eV}^{-1}.
\end{equation}
These shifts perturb the one-pion-exchange potential~\cite{Ubaldi:2008nf}, thereby affecting the relevant nuclear binding energies. For tritium, this perturbation modifies the triton binding energy, which we express as
\begin{equation}
B_3(\phi)=8.1\times10^{6}~\mathrm{eV}
+\Bigg[-1.06\times10^{-10}(\lambda_{12}\phi)^2-2.26\times10^{-12}(\lambda_{13}\phi)^2+1.93\times10^{-12}(\lambda_{23}\phi)^2\Bigg]\mathrm{eV}^{-1}.
\end{equation}
The shifts in both the proton--neutron mass difference and the triton binding energy induce oscillations in the tritium $\beta$-decay rate. These oscillations are encoded in the fractional deviation of the phase-space integral:
\begin{equation}\label{eq:IPhi}
\frac{\delta I^\beta}{I^\beta}
=\Bigg[-1.32\times10^{-12}(\lambda_{12}\phi)^2-2.81\times10^{-14}(\lambda_{13}\phi)^2-3.29\times10^{-17}(\lambda_{23}\phi)^2\Bigg]\mathrm{eV}^{-2}.
\end{equation}
Combining with Eq.~\eqref{eq:betaGammChange}, we obtain
\begin{equation}\label{eq:I0Phi}
\begin{aligned}
I_0
=&\Bigg[-1.32\times10^{-12}(\lambda_{12}\phi)^2-2.81\times10^{-14}(\lambda_{13}\phi)^2-3.29\times10^{-17}(\lambda_{23}\phi)^2\Bigg]\mathrm{eV}^{-2}\\
&+\left[-5.20\times10^{-9}\lambda_{12}\phi-1.83\times10^{-12}\lambda_{13}\phi\right]\mathrm{eV}^{-1}.
\end{aligned}
\end{equation}
Here $I_0$ denotes the fractional deviation of the tritium $\beta$-decay rate from its standard value; $\Gamma_0$ is the decay rate in the absence of ULDM. This oscillatory behavior is central to identifying ULDM through its impact on nuclear processes such as tritium decay.

To incorporate constraints from tritium decay measurements, following the analogous analysis in Ref.~\cite{Zhang:2023lem}, we model the discrete Fourier transform (DFT) coefficient at a trial frequency bin $p$ as
\begin{equation}
\tilde{d}_p=\tilde{n}_p+\tilde{I}_p,
\end{equation}
where $\tilde{n}_p$ represents complex Gaussian noise with variance $N\sigma^2$ (with $N$ the number of data points), and $\tilde{I}_p$ is the DFT of the deterministic modulation $I(\phi)$, taking $\phi(t)=\Phi_0\cos(m_\phi t+\phi_0)$. After marginalizing over the unknown signal phase, the normalized modulus $A\equiv|\tilde{d}_p|/\sqrt{N\sigma^2}$ follows a Rician distribution (equivalently, a noncentral $\chi_2^2$ distribution),
\begin{equation}
\mathcal{L}\!\left(A\,\middle|\,A_s\right)=2A\,e^{-A^2-A_s^2}I_0\!\left(2AA_s\right),
\end{equation}
where $A_s=|\tilde{I}_p|/\sqrt{N\sigma^2}$. In the stochastic regime, where $\tau_{\rm coh}\ll\tau_\phi$, the field amplitude $\Phi_0$ is  distributed with scale $\bar\phi=\sqrt{2\rho_{\mathrm{DM}}}/m_\phi$, and the likelihood is marginalized over $\Phi_0$:
\begin{equation}
\mathcal{L}_{\rm marg}(A\,|\,\bar A_s)=\int_0^\infty \mathcal{L}\!\big(A\,|\,A_s(\Phi_0)\big)\,
\underbrace{\frac{2\Phi_0}{\bar{\phi}^2}e^{-\Phi_0^2/\bar\phi^2}}_{p(\Phi_0)}\,d\Phi_0,
\end{equation}
where $p(\Phi_0)$ is the Rayleigh distribution for $\Phi_0$. This procedure accounts for the uncertainty in the field amplitude and determines the signal strength as a function of the coupling parameters.

\begin{figure}[htbp]
    \begin{subfigure}[b]{0.48\linewidth}
        \centering
        \includegraphics[width=\linewidth]{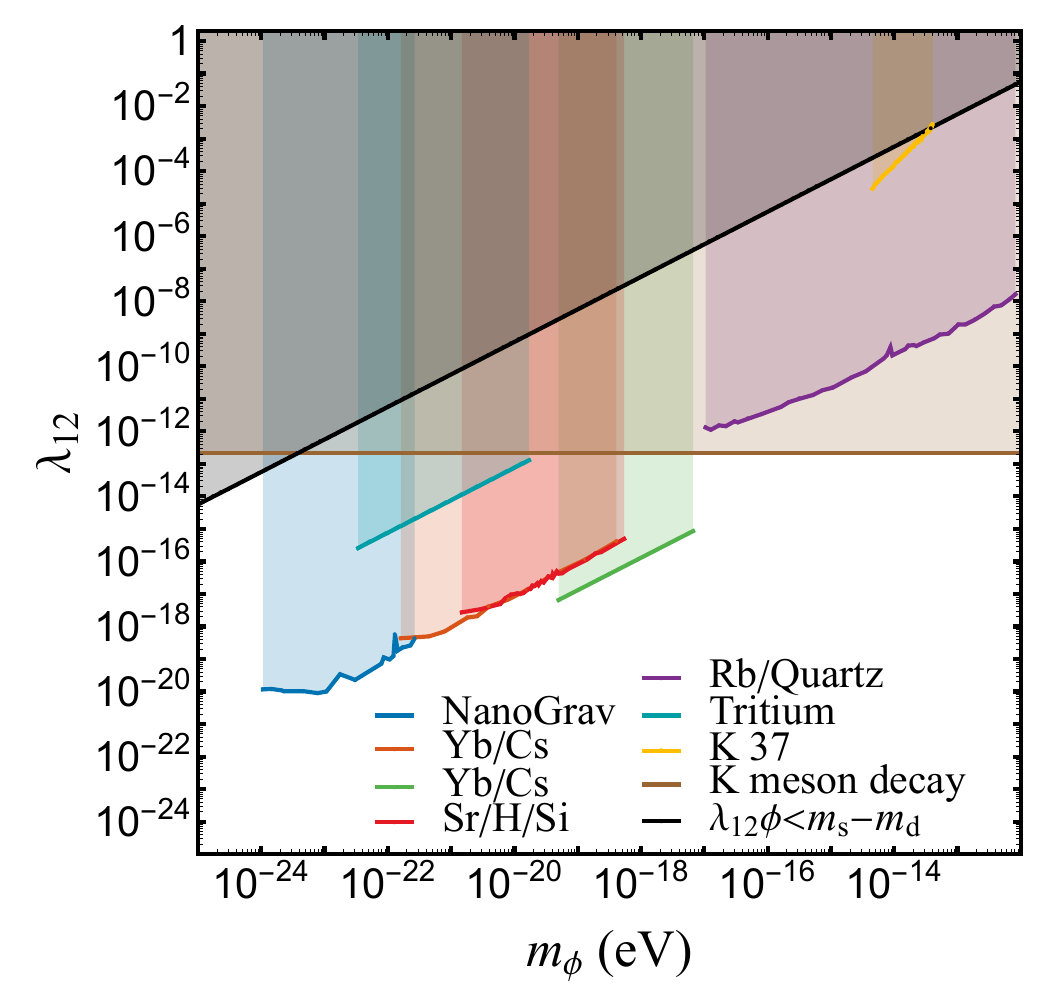}
        \caption{Constraint on $\lambda_{12}$}
        \label{fig:lambda12}
    \end{subfigure}%
    \hfill
    \begin{subfigure}[b]{0.48\linewidth}
        \centering
        \includegraphics[width=\linewidth]{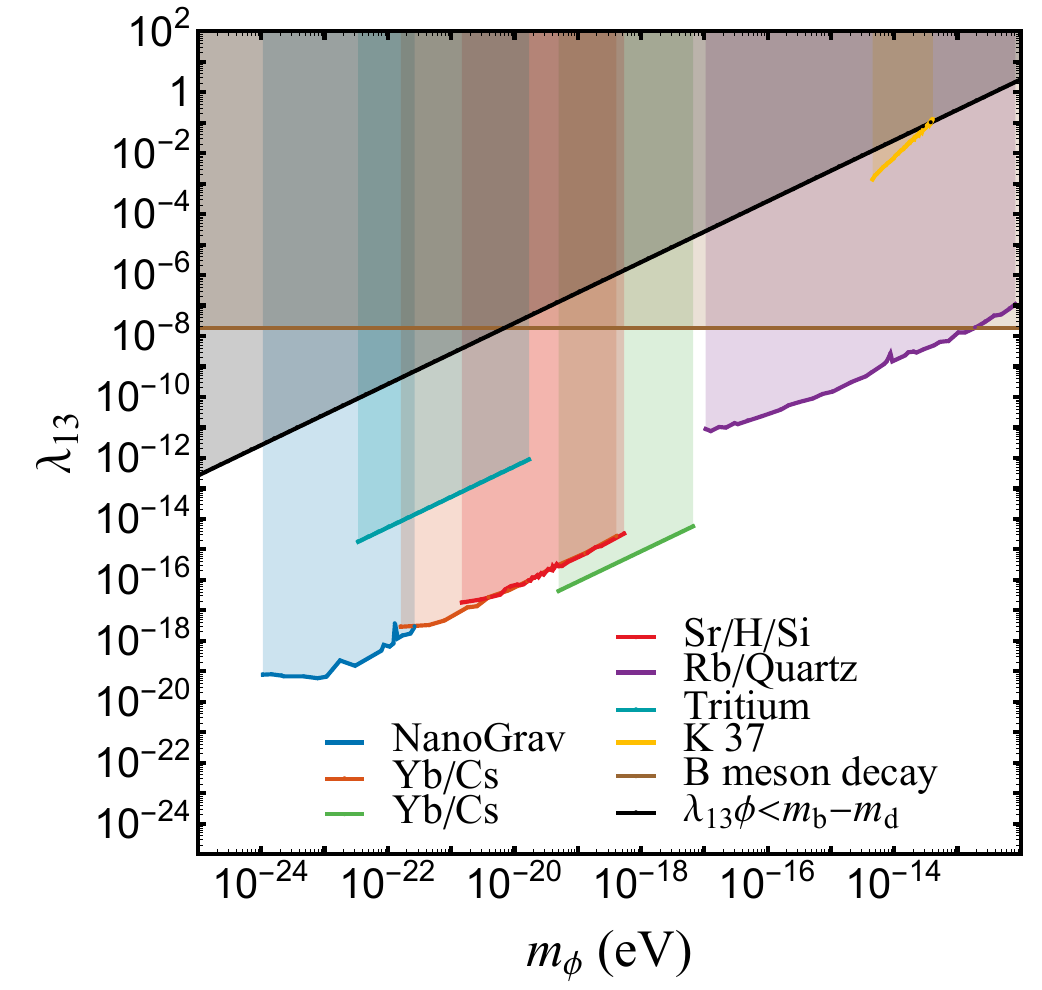}
        \caption{Constraint on $\lambda_{13}$}
        \label{fig:lambda13}
    \end{subfigure}%
    \hfill
    \begin{subfigure}[b]{0.48\linewidth}
        \centering
        \includegraphics[width=\linewidth]{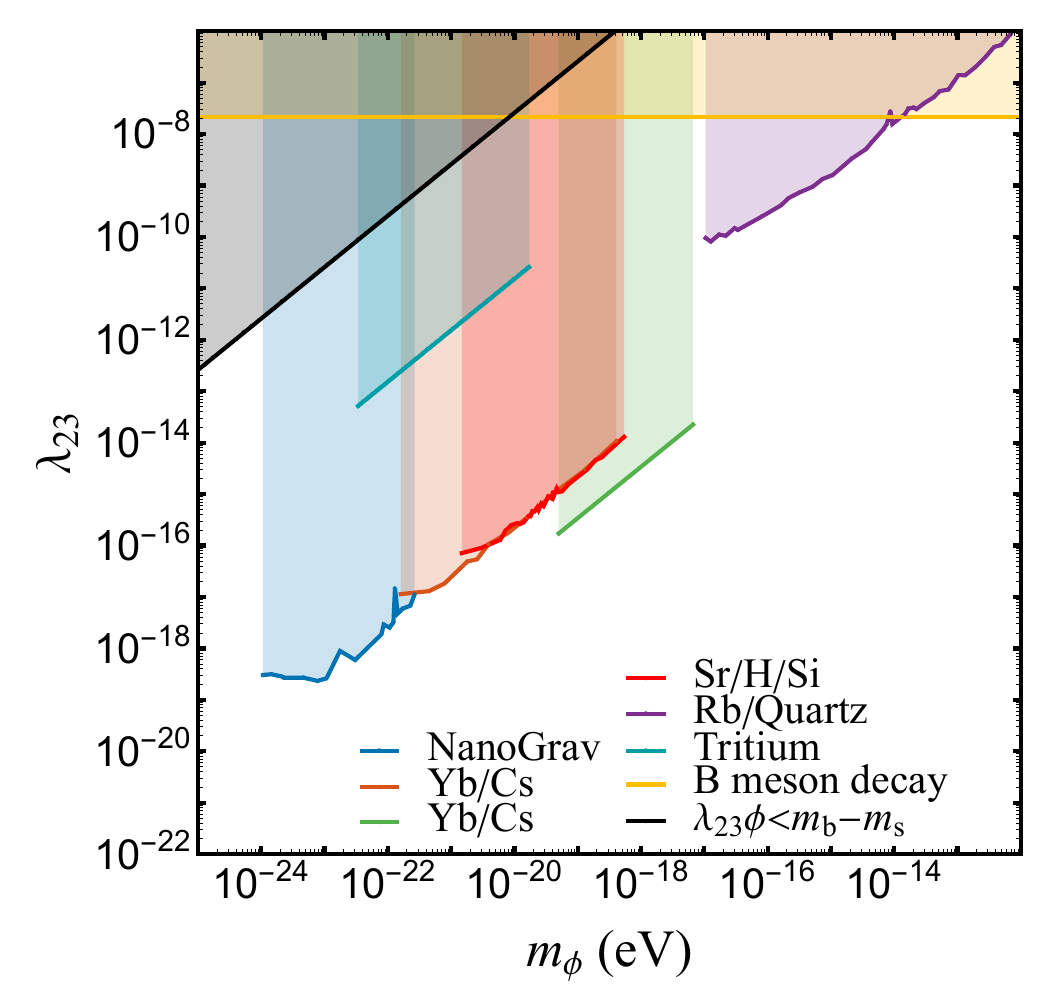}
        \caption{Constraint on $\lambda_{23}$}
        \label{fig:Lambda23}
    \end{subfigure}
    \captionsetup{justification=raggedright}
    \caption{Comparison of the constraints on ULDM derived from NANOGrav~\cite{NANOGrav:2023hvm} and Yb/Cs clocks~\cite{Sherrill:2023zah,Kobayashi:2022vsf} as well as Sr/H/Si clocks~\cite{Kennedy:2020bac}, tritium beta decay~\cite{tritiumdecay}, ${}^{37}\mathrm{K}$ decay~\cite{Shidling:2014ura}, meson decay~\cite{Belle-II:2023esi,NA62:2020xlg,Belle:2017oht}, and $\mathrm{Rb/Quartz}$ clock~\cite{Zhang:2022ewz}. The black line represents $|\lambda_{ij}\bar\phi/(m_i-m_j)|$.}
    \label{fig:Combined-constraint}
\end{figure}

Next, we compute the detection threshold and exclusion limits. The $95\%$ detection threshold $A^{\mathrm{th}}$ is determined by the false-positive rate $\alpha=0.05$:
\begin{equation}
\alpha=\int_{A^{\mathrm{th}}}^{\infty}\mathcal{L}(A\,|\,0)\,dA.
\end{equation}
The miss probability at signal strength $\bar A_s$ is
\begin{equation}
\beta(\bar A_s)=\int_0^{A^{\mathrm{th}}}\mathcal{L}_{\rm marg}(A\,|\,\bar A_s)\,dA.
\end{equation}
In this work, we define the exclusion coupling by the equal-error criterion at $\alpha=0.05$, i.e.\ we solve $\alpha=\beta(\bar A_s)$ for $\bar A_s$. Using $I(\phi)$ from Eq.~\eqref{eq:IPhi} and evaluating the DFT at the bin matched to $2m_\phi$ for the second-order terms yields
\begin{equation}
\begin{aligned}
A_s=\frac{|\tilde I_p|}{\sqrt{N\sigma^2}}
=&\Bigg|\Bigg(
-6.6\times10^{-13}(\lambda_{12}\phi)^2-1.41\times10^{-14}(\lambda_{13}\phi)^2-1.65\times10^{-17}(\lambda_{23}\phi)^2
\Bigg)\frac{\sqrt{N}}{\sigma}~\mathrm{eV}^{-2}\\
&+\left[-5.20\times10^{-9}\lambda_{12}\phi-1.83\times10^{-12}\lambda_{13}\phi\right]\frac{\sqrt{N}}{\sigma}\mathrm{eV}^{-1}\Bigg|.
\end{aligned}
\end{equation}
Substituting this $A_s$ into $\beta(\bar A_s)$ and imposing $\alpha=\beta$ at $\alpha=0.05$ leads to
\begin{equation}
\begin{aligned}
&\left(\frac{1.73\times10^{-14}}{m_\phi^2}\lambda_{12}^2+\frac{2.98\times10^{-28}}{m_\phi^2}\lambda_{13}^2
+\frac{4.29\times10^{-19}}{m_\phi^2}\lambda_{23}^2\right)\mathrm{eV}^2\lesssim 1,
\end{aligned}
\end{equation}
which holds in the coherence-dominated regime $\tau_[\rm coh]\ll\tau_{\rm coh}$, where $m_\phi\in\left[3.4\times10^{-23},\,1.7\times10^{-20}\right]\mathrm{eV}$. The strong constraint is driven by the second-order modification of the phase-space integral. This constraint is visualized in Fig.~\ref{fig:Combined-constraint}. Specifically, we obtain $95\%$ CL upper limits $\lambda_{12}\lesssim 10^{-16}$ and $\lambda_{13}\lesssim 10^{-15}$ near $m_\phi=3.4\times10^{-23}\,\mathrm{eV}$, placing stringent bounds on the interaction strength between ULDM and the SM.

\subsection{Constraints from Quark Mass Measurements}
As shown in Eq.~\eqref{eq:quark-mass-change}, the ULDM background shifts the down-type quark masses by an amount proportional to $\phi^2$. We therefore interpret the experimental uncertainties on quark mass determinations as upper bounds on any additional DM-induced contributions.

We adopt the PDG $\overline{\rm MS}$ masses and quoted $1\sigma$ uncertainties~\cite{ParticleDataGroup:2024cfk}:
$m_d=(4.70\pm0.07)\,{\rm MeV}$,
$m_s=(93.5\pm0.8)\,{\rm MeV}$,
$m_b=(4.183\pm0.007)\,{\rm GeV}$,
and require that the ULDM-induced shift does not exceed these uncertainties.

For an oscillating ULDM condensate, using Eq.~\eqref{eq:quark-mass-change}, the time-averaged mass shift
$\delta m_q\equiv \langle m_q'-m_q\rangle \propto \langle\phi^2\rangle$ is quadratic in the couplings.
In the one-coupling-at-a-time approximation, we derive the strongest bound by requiring that the corresponding ULDM-induced mass shift not exceed the quoted $1\sigma$ uncertainty of the most sensitive quark-mass determination ($\sigma(m_q)$). This gives
\begin{equation}
\lambda_{ij}\ \lesssim\ m_\phi\left(\frac{\sigma(m_q)\,|m_i-m_j|}{\rho_{\rm DM}}\right)^{1/2},
\end{equation}
where $q=d$ for $\lambda_{12}$ and $\lambda_{13}$, while $q=s$ for $\lambda_{23}$.
With $\rho_{\rm DM}=0.4~{\rm GeV/cm^{3}}$, we obtain
\begin{equation}
\begin{aligned}
&\lambda_{12}\lesssim 1.4\times10^{-5}\left(\frac{m_\phi}{10^{-14}\,{\rm eV}}\right)\left(\frac{0.4~{\rm GeV\,cm^{-3}}}{\rho_{\rm DM}}\right)^{1/2},\\
&\lambda_{13}\lesssim 9.8\times10^{-5}\left(\frac{m_\phi}{10^{-14}\,{\rm eV}}\right)\left(\frac{0.4~{\rm GeV\,cm^{-3}}}{\rho_{\rm DM}}\right)^{1/2},\\
&\lambda_{23}\lesssim 3.3\times10^{-4}\left(\frac{m_\phi}{10^{-14}\,{\rm eV}}\right)\left(\frac{0.4~{\rm GeV\,cm^{-3}}}{\rho_{\rm DM}}\right)^{1/2}.
\end{aligned}
\end{equation}

\subsection{Atomic Clocks} 
ULDM induces oscillations in the quark masses, as shown in Eq.~\eqref{eq:quark-mass-change}, which in turn lead to periodic variations in fundamental constants such as the proton mass, electron mass, and the fine-structure constant. Such variations can be probed with precision atomic clocks, which measure atomic transition frequencies. By comparing clocks with different sensitivities to these constants, one can isolate the effects of ULDM on atomic energy levels, providing a powerful method to constrain ULDM~\cite{Elder:2025tue,Zhang:2022ewz}.

A variety of existing atomic-clock measurements have already set stringent limits on ULDM-induced oscillations. These experiments constrain variations of fundamental quantities---including the proton and electron masses, (hyper-)fine-structure parameters, and related combinations---through high-precision frequency comparisons across different atomic species. For a comprehensive review of this subject, see Refs.~\cite{Filzinger:2023qqh,Fuchs:2024xvc,2022PhRvL.129x1301K,Madge:2024aot,Elder:2025tue,Flambaum:2004tm}. In particular, Ref.~\cite{Elder:2025tue} provides a systematic discussion of ULDM searches with next-generation atomic and nuclear clocks, both theoretically and experimentally. Here we focus on recent frequency-comparison measurements involving Yb/Cs~\cite{Sherrill:2023zah,Kobayashi:2022vsf}, Sr/H/Si~\cite{Kennedy:2020bac}, and Rb/Quartz~\cite{Zhang:2022ewz}, which currently provide relatively stringent bounds on ULDM-induced oscillations in atomic energy levels.

These clock-comparison measurements yield upper limits on the fractional change in the clock frequency ratio, $\delta f/f$. Using sensitivity coefficients~\cite{Kozlov:2018qid}, the results can be translated into bounds on variations of fundamental constants, including the fine-structure constant $\alpha$, the electron mass $m_e$, the QCD scale $\Lambda_{\rm QCD}$, and quark masses, among others. For the present ULDM--quark interaction model, the corresponding constraints follow from relating the ULDM-induced oscillations to variations of fundamental constants, in particular through the induced nucleon couplings in Eq.~\eqref{eq:eff_p_n}. These nucleon couplings enter clock observables via
\begin{equation}
\frac{\delta f}{f}=\sum_i K_i\,\frac{\delta X_i}{X_i},
\end{equation}
where $f$ denotes the frequency ratio of the two compared clock transitions, and $X_i$ are the underlying dimensionless fundamental constants (or appropriate dimensionless combinations) to which this ratio is sensitive, e.g., $X_i\in\{\alpha,\ m_e/\Lambda_{\rm QCD},\ m_q/\Lambda_{\rm QCD},\ \ldots\}$. The coefficients $K_i\equiv \partial\ln f/\partial\ln X_i$ quantify the logarithmic sensitivity of the frequency ratio to each $X_i$, while $\delta X_i/X_i$ are the corresponding fractional variations induced by the ULDM background~\cite{Kozlov:2018qid}.

In Fig.~\ref{fig:Combined-constraint}, we present the 95\% CL upper bounds derived from the Yb/Cs, Sr/H/Si, and Rb/Quartz clock comparisons. These bounds are shown in different colors and correspond to constraints on the ULDM coupling parameters $\lambda_{12}$, $\lambda_{13}$, and $\lambda_{23}$.
In addition, as shown in Ref.~\cite{NANOGrav:2023hvm}, the 15-year NANOGrav pulsar-timing dataset also constrains ULDM. A direct reanalysis of pulsar timing residuals is applied to our model, and the resulting bounds are included in Fig.~\ref{fig:Combined-constraint}. The dominant contributions to the timing residuals arise from pulsar spin fluctuations~\cite{Kaplan:2022lmz} and reference-clock shifts~\cite{Kaplan:2022lmz, Graham:2015ifn}. For the former, nucleon-mass oscillations modify the pulsar moment of inertia and hence induce spin fluctuations. For the latter, PTA timing residuals---which rely on cesium atomic clocks~\cite{mccarthy2018time}---are affected by ULDM-induced shifts in clock frequencies. The strongest NANOGrav constraints are
$\lambda_{12}<1.16\times 10^{-20}$, $\lambda_{13}<7.73\times10^{-20}$, and
$\lambda_{23}<3.05\times10^{-19}$ around $m_\phi\sim 10^{-24}~\mathrm{eV}$.
For heavier dark-matter masses, Rb/Quartz yields
$\lambda_{12}<1.63\times 10^{-8}$, $\lambda_{13}<1.11\times 10^{-7}$, and
$\lambda_{23}<1.19\times 10^{-6}$ at $m_\phi = 8.13\times 10^{-11}~\mathrm{eV}$.

Beyond the clock systems considered here, other platforms---including single-ion Al$^+$ and Hg$^+$ optical clocks~\cite{Rosenband:2008qgq,Bize:2002vap}, Yb$^+$ octupole and quadrupole clocks~\cite{Huntemann:2012zz}, Dy-based optical clocks, and the proposed $^{229}$Th nuclear clock~\cite{vonderWense:2020bbs}---also merit further study. Their analyses follow an analogous methodology, and we therefore do not discuss them in detail here.

\subsection{Meson Mixing from a ULDM Background Field}
\label{sec:kaonmix_ckm}

In the presence of a ULDM background, the charged-current interaction is governed by the time-dependent CKM matrix $V'_{\rm CKM}$ in Eq.~\eqref{eq:CKM-matrix}. As a consequence, neutral-meson mass splittings acquire a periodic modulation with angular frequency $m_\phi$.

We define the quark-mass splittings
\begin{equation}
\Delta_{sd}\equiv m_s-m_d,\qquad
\Delta_{bd}\equiv m_b-m_d,\qquad
\Delta_{bs}\equiv m_b-m_s.
\end{equation}

For numerical estimates, we use PDG central values for CKM magnitudes, quark masses, and neutral-meson mass splittings, rounded to the precision required here~\cite{ParticleDataGroup:2024cfk}:
$|V_{cd}|=0.22487$, $|V_{cs}|=0.97349$, $|V_{cb}|=0.04183$, $|V_{td}|=8.6\times 10^{-3}$, $|V_{ts}|=41.5\times 10^{-3}$, $|V_{tb}|\simeq 1$, $m_d=4.7~{\rm MeV}$, $m_s=93~{\rm MeV}$, $m_b=4.18~{\rm GeV}$, which imply $\Delta_{sd}=88.3~{\rm MeV}$, $\Delta_{bd}=4.175~{\rm GeV}$, and $\Delta_{bs}=4.087~{\rm GeV}$. For the experimental mass splittings and their relative uncertainties, we take $\Delta m_K=0.5293\times 10^{-2}\,{\rm ps}^{-1}$ and $\delta_K=1.70\times 10^{-3}$, $\Delta m_d=0.5069~{\rm ps}^{-1}$ and $\delta_{B_d}=3.75\times 10^{-3}$, $\Delta m_s=17.765~{\rm ps}^{-1}$ and $\delta_{B_s}=3.38\times 10^{-4}$. We parametrize the local ULDM density by $\rho_{\rm DM}$, and use $0.4~{\rm GeV\,cm^{-3}}$ as a reference value~\cite{deSalas:2020hbh}.

For an estimate of the modulation in $\Delta m_K$, we retain the dominant charm contribution to the real part of the short-distance amplitude~\cite{Buras:2014maa}. The fractional modulation is
\begin{equation}
\frac{\delta(\Delta m_K)(t)}{\Delta m_K}\simeq
2\,{\rm Re}\!\left(\frac{\delta\lambda_c^K(t)}{\lambda_c^{K,{\rm SM}}}\right),
\qquad
\lambda_c^{K,{\rm SM}}=V_{cs}^*V_{cd}.
\label{eq:frac_deltaDeltamK}
\end{equation}
At linear order in the ULDM-induced CKM shifts from Eq.~\eqref{eq:CKM-matrix}, one finds
\begin{equation}
\frac{\delta\lambda_c^K}{\lambda_c^{K,{\rm SM}}}
=\phi\left[
\frac{\lambda_{12}}{\Delta_{sd}}\,
\frac{|V_{cd}|^2-|V_{cs}|^2}{V_{cs}^*V_{cd}}
-\frac{\lambda_{23}}{\Delta_{bs}}\,
\frac{V_{cb}^*}{V_{cs}^*}
-\frac{\lambda_{13}}{\Delta_{bd}}\,
\frac{V_{cb}}{V_{cd}}
\right].
\label{eq:delta_lambdac_over_lambdac}
\end{equation}

We take the ULDM background as in Eq.~\eqref{DM-eq}. Let $\delta_K$ denote the allowed fractional modulation amplitude of $\Delta m_K$. A sufficient condition is
\begin{equation}
2\,|\bar{\phi}|\left[
\frac{|\lambda_{12}|}{\Delta_{sd}}\,
\frac{\big||V_{cd}|^2-|V_{cs}|^2\big|}{|V_{cs}||V_{cd}|}
+\frac{|\lambda_{23}|}{\Delta_{bs}}\,
\frac{|V_{cb}|}{|V_{cs}|}
+\frac{|\lambda_{13}|}{\Delta_{bd}}\,
\frac{|V_{cb}|}{|V_{cd}|}
\right]\le \delta_K.
\label{eq:combined_bound_lambdas}
\end{equation}
This yields the single-coupling limits
\begin{align}
|\lambda_{12}|&\le
\frac{\delta_K}{2|\bar\phi|}\,
\Delta_{sd}\,
\frac{|V_{cs}||V_{cd}|}{\big||V_{cd}|^2-|V_{cs}|^2\big|},
\label{eq:bound_lambda12}
\\
|\lambda_{13}|&\le
\frac{\delta_K}{2|\bar\phi|}\,
\Delta_{bd}\,
\frac{|V_{cd}|}{|V_{cb}|},
\label{eq:bound_lambda13}
\\
|\lambda_{23}|&\le
\frac{\delta_K}{2|\bar\phi|}\,
\Delta_{bs}\,
\frac{|V_{cs}|}{|V_{cb}|}.
\label{eq:bound_lambda23}
\end{align}

With the numerical inputs, we obtain
\begin{align}
|\lambda_{12}|&\leq
7.4\times 10^{-16}\,
\Bigl[\frac{m_\phi}{10^{-22}~{\rm eV}}\Bigr]
\Bigl[\frac{0.4~{\rm GeV\,cm^{-3}}}{\rho_{\rm DM}}\Bigr]^{1/2},
\label{eq:bound_lambda12_numeric}
\\
|\lambda_{13}|&\leq
7.7\times 10^{-13}\,
\Bigl[\frac{m_\phi}{10^{-22}~{\rm eV}}\Bigr]
\Bigl[\frac{0.4~{\rm GeV\,cm^{-3}}}{\rho_{\rm DM}}\Bigr]^{1/2},
\label{eq:bound_lambda13_numeric}
\\
|\lambda_{23}|&\leq
3.3\times 10^{-12}\,
\Bigl[\frac{m_\phi}{10^{-22}~{\rm eV}}\Bigr]
\Bigl[\frac{0.4~{\rm GeV\,cm^{-3}}}{\rho_{\rm DM}}\Bigr]^{1/2}.
\label{eq:bound_lambda23_numeric}
\end{align}

The same CKM modulation also induces constraints from $B_d^0-\bar B_d^0$ and $B_s^0-\bar B_s^0$ mixing. The short-distance amplitude is dominated by the top contribution, leading to the estimates
\begin{equation}
\frac{\delta(\Delta m_d)(t)}{\Delta m_d}
\simeq
2\,{\rm Re}\!\left(\frac{\delta\lambda_t^{(d)}(t)}{\lambda_{t,\rm SM}^{(d)}}\right),
\qquad
\lambda_{t,\rm SM}^{(d)}=V_{td}^*V_{tb},
\label{eq:frac_delta_Dmd}
\end{equation}
\begin{equation}
\frac{\delta(\Delta m_s)(t)}{\Delta m_s}
\simeq
2\,{\rm Re}\!\left(\frac{\delta\lambda_t^{(s)}(t)}{\lambda_{t,\rm SM}^{(s)}}\right),
\qquad
\lambda_{t,\rm SM}^{(s)}=V_{ts}^*V_{tb}.
\label{eq:frac_delta_Dms}
\end{equation}

Keeping the parametrically dominant terms in the single-coupling limits, one obtains
\begin{equation}\label{eq:bound_lambda23_Bs}
|\lambda_{13}|\le
\frac{\delta_{B_d}}{2|\bar\phi|}\,
\Delta_{bd}\,
\frac{|V_{td}|}{|V_{tb}|},~~
|\lambda_{23}|\le
\frac{\delta_{B_s}}{2|\bar\phi|}\,
\Delta_{bs}\,
\frac{|V_{ts}|}{|V_{tb}|}.
\end{equation}
Numerically,
\begin{align}
|\lambda_{13}|&\le
2.7\times 10^{-15}\,
\Bigl[\frac{m_\phi}{10^{-22}~{\rm eV}}\Bigr]
\Bigl[\frac{0.4~{\rm GeV\,cm^{-3}}}{\rho_{\rm DM}}\Bigr]^{1/2},
\label{eq:bound_lambda13_Bmix_numeric}
\\
|\lambda_{23}|&\le
1.2\times 10^{-15}\,
\Bigl[\frac{m_\phi}{10^{-22}~{\rm eV}}\Bigr]
\Bigl[\frac{0.4~{\rm GeV\,cm^{-3}}}{\rho_{\rm DM}}\Bigr]^{1/2}.
\label{eq:bound_lambda23_Bmix_numeric}
\end{align}
These bounds are stronger than the kaon-mixing limits for $\lambda_{13}$ and $\lambda_{23}$, while kaon mixing provides the leading constraint on $\lambda_{12}$.

\section{Constraints on the ULDM Field as a Quantum Particle}
\label{sec:particle}
As discussed in Sec.~\ref{sec:model}, the scalar field $\phi$ can also be treated as a quantum field describing particle excitations. In this section, we use the interaction terms and the fifth-force expressions derived above to obtain constraints from quark flavor-violating decays and searches for exotic fifth forces.

\subsection{Quark-flavor Violating Decays}
The off-diagonal interaction induces quark flavor-violating decays, which at the hadronic level lead to flavor-changing processes such as
\begin{equation}
B\to K\phi\quad \text{and} \quad K\to \pi \phi.
\end{equation}
Since $\phi$ is invisible to detectors, these processes contribute to the experimental signatures $B\to K+\text{Inv}$ and $K\to \pi+\text{Inv}$. Measurements of the corresponding branching ratios therefore constrain the coupling strengths $\lambda_{ij}$.

Following Ref.~\cite{Kachanovich:2020yhi}, the branching ratios for the two-body decays $B\to K\phi$ and $K\to \pi\phi$ can be written as
\begin{equation}\label{eq:BR_BtoKphi}
\begin{aligned}
\mathcal{B}(B\to K\phi)
&=\frac{\tau_B}{8\pi\,m_B^2}\,\lambda_{23}^2
\left(\frac{m_B^2-m_K^2}{m_b-m_s}\right)^{\!2}
f_{0,BK}(m_{\phi}^2)^{2}\,
\frac{\lambda^{1/2}(m_B^2,m_K^2,m_{\phi}^2)}{2m_B},\\
\mathcal{B}(K\to \pi\phi)
&=\frac{\tau_K}{8\pi\,m_K^2}\,\lambda_{12}^2
\left(\frac{m_K^2-m_\pi^2}{m_s-m_d}\right)^{\!2}
f_{0,K\pi}(m_{\phi}^2)^{2}\,
\frac{\lambda^{1/2}(m_K^2,m_\pi^2,m_{\phi}^2)}{2m_K},\\
\mathcal{B}(B\to \pi\phi)&=\frac{\tau_B}{8\pi m_B^2}\lambda_{13}^2
\left(\frac{m_B^2-m_\pi^2}{m_b-m_d}\right)^2f_{0,B\pi}(m_\phi^2)^2
\frac{\lambda^{1/2}(m_B^2,m_\pi^2,m_\phi^2)}{2m_B},
\end{aligned}
\end{equation}
where $\tau_B$ ($\tau_K$) and $m_B$ ($m_K$) denote the lifetime and mass of the $B$ ($K$) meson, respectively, and $\lambda(a,b,c)=a^2+b^2+c^2-2(ab+ac+bc)$ is the Källén function. The scalar form factors $f_{0,BK}(q^2)$, $f_{0,K\pi}(q^2)$, and $f_{0,B\pi}(q^2)$ are defined by
\begin{equation}
\label{eq:scalarFF}
\begin{aligned}
\langle K|\bar{s}b|B\rangle
&= \frac{m_B^2-m_K^2}{m_b-m_s}\,f_{0,BK}(q^2),
\qquad q=p_B-p_K,\\
\langle\pi|\bar{s}d|K\rangle &=\frac{m_K^2-m_\pi^2}{m_s-m_d}f_{0,K\pi}(q^2),\qquad
q = p_K-p_\pi,\\
\langle\pi|\bar{d} b|B\rangle &=\frac{m_B^2-m_\pi^2}{m_b-m_d}f_{0,B\pi}(q^2),\qquad
q = p_B-p_\pi,
\end{aligned}
\end{equation}
with lattice QCD values $f_{0,BK}(q^2)=0.309$, $f_{0,K\pi}(q^2)\approx1$, and $f_{0,B\pi}(q^2)= 0.26$ from Refs.~\cite{Bailey:2015dka,Carrasco:2016kpy,Gubernari:2018wyi}.

For conservative estimates, we use the following representative $90\%$ CL bounds on final states with missing energy, based on Refs.~\cite{Belle-II:2023esi,NA62:2020xlg,Belle:2017oht}:
\begin{equation}
\begin{aligned}
&\mathcal{B}(B^+\to K^+ + \text{Inv}) \;\lesssim\; 1.9\times 10^{-5}\quad(90\%~\text{CL}),\\
&\mathcal{B}(K^+\to \pi^+ + \text{Inv}) \;\lesssim\; 2\times 
10^{-10}\quad(90\%~\text{CL}),\\
&\mathcal{B}(B^+\to \pi^+ + \text{Inv}) \;\lesssim\; 8\times 10^{-6}\quad(90\%~\text{CL}).\\
\end{aligned}
\end{equation}
Combining these limits with Eq.~\eqref{eq:BR_BtoKphi} yields bounds on the couplings:
\begin{equation}\label{eq:bound_lambda23}
\begin{aligned}
&\lambda_{23}\lesssim
\left[
\frac{8\pi\,m_B^2}{\tau_{B^+}}
\left(\frac{m_b-m_s}{m_B^2-m_K^2}\right)^{\!2}
\frac{2m_B}{\lambda^{1/2}(m_B^2,m_K^2,m_{\phi}^2)}\;
\frac{\mathcal{B}(B^+\to K^+ + \text{Inv})}{f_{0,B K}(m_\phi^2)^2}
\right]^{1/2},\\
&\lambda_{12}\lesssim
\left[
\frac{8\pi\,m_K^2}{\tau_{K^+}}
\left(\frac{m_s-m_d}{m_K^2-m_\pi^2}\right)^{\!2}
\frac{2m_K}{\lambda^{1/2}(m_K^2,m_\pi^2,m_{\phi}^2)}\;
\frac{\mathcal{B}(K^+\to \pi^+ + \text{Inv})}{f_{0,K\pi}(m_\phi^2)^2}
\right]^{1/2},\\
&\lambda_{13}\lesssim
\left[
\frac{8\pi\,m_B^2}{\tau_{B^+}}
\left(\frac{m_b-m_d}{m_B^2-m_\pi^2}\right)^{\!2}
\frac{2m_B}{\lambda^{1/2}(m_B^2,m_\pi^2,m_{\phi}^2)}\;
\frac{\mathcal{B}(B^+\to \pi^+ + \text{Inv})}{f_{0,B\pi}(m_\phi^2)^2}
\right]^{1/2}.\\
\end{aligned}
\end{equation}
Since $\phi$ serves as a ULDM field, its mass can be omitted in deriving the constraint for $m_\phi\lesssim {\rm eV}$, which yields
\begin{equation}
\label{eq:bound_Lambda23}\lambda_{23}<2.1\times 10^{-8},~~
\lambda_{12}<2.1\times 10^{-13},~~
\lambda_{13}<1.8\times 10^{-8}.
\end{equation}
The corresponding results are shown in Fig.~\ref{fig:Combined-constraint}.

\subsection{Constraint from Meson Oscillations}

Consider the interaction $\mathcal L \supset -\lambda_{12}\,\phi\,(\bar s d+\bar d s)$ in Eq.~\eqref{eq:eff_lag}. Tree-level $\phi$ exchange generates a bilocal $\Delta S=2$ interaction. Adopting the equal-time approximation for the bilocal quark densities and performing the time-difference integral of the $\phi$ propagator yields a static Yukawa kernel. We parametrize the resulting momentum weighting by a single effective scale $q_{\rm eff}(K)$, which defines
\begin{equation}\label{eq:HeffK_local_short}
H_{\rm eff}^{\Delta S=2}\simeq
\frac{1}{2}\lambda_{12}^2\,\overline{\Delta}_K(m_\phi)\,
\bigl[ O_2+O_2'+2O_4 \bigr],
\qquad
\overline{\Delta}_K(m_\phi)\equiv \frac{1}{q_{\rm eff}(K)^2+m_\phi^2},
\end{equation}
where $O_i$ are the local $\Delta S=2$ operators in the SUSY basis~\cite{UTfit:2007eik}. For the kaon we take $q_{\rm eff}(K)=0.3759~{\rm GeV}$ from the light-front quark model (LFQM) variational fit of Ref.~\cite{Choi:2015ywa}.

We evaluate the required matrix elements at $\mu=3~{\rm GeV}$ in $\overline{\rm MS}$ using lattice bag parameters~\cite{Boyle:2024gge} and  FLAG~\cite{FlavourLatticeAveragingGroupFLAG:2024oxs},
\begin{equation}\label{eq:O2O4_numbers_short}
\langle \bar K^0|O_2|K^0\rangle \simeq -0.153~{\rm GeV}^4,
\qquad
\langle \bar K^0|O_4|K^0\rangle \simeq 0.344~{\rm GeV}^4.
\end{equation}
Parity invariance implies $\langle O_2'\rangle=\langle O_2\rangle$ for pseudoscalar external states. The dispersive contribution is
\begin{equation}\label{eq:DmKphi_short}
\Delta m_K^{\phi}\simeq 2\,{\rm Re}\,M_{12}^{\phi},
\qquad
M_{12}^{\phi}=
\frac{\lambda_{12}^2\,\overline{\Delta}_K(m_\phi)}{4m_K}\,
\bigl[2\langle O_2\rangle+2\langle O_4\rangle\bigr]
\simeq
\lambda_{12}^2\,\overline{\Delta}_K(m_\phi)\times 0.192~{\rm GeV}^3,
\end{equation}
with $m_K$ and $(\Delta m_K)_{\rm exp}$ taken from the PDG~\cite{ParticleDataGroup:2024cfk}. In the ultralight regime $m_\phi\ll q_{\rm eff}(K)$,
\begin{equation}
\Delta m_K^{\phi}\simeq (2.72~{\rm GeV})\,\lambda_{12}^2,
\qquad
|\Delta m_K^{\phi}|\le (\Delta m_K)_{\rm exp}
\ \Rightarrow\
\lambda_{12}\lesssim 3.5\times 10^{-8}.
\label{eq:lambda12_bound_short}
\end{equation}

The interactions
\begin{equation}
\mathcal{L}\ \supset\ -\lambda_{13}\,\phi\,(\bar b d+\bar d b)\ -\ \lambda_{23}\,\phi\,(\bar b s+\bar s b)
\end{equation}
induce $\Delta B=2$ transitions via the same mechanism. The static-kernel result can be expressed in terms of the HPQCD reduced matrix elements $R_{2,4}^{(q)}$ at $\mu=m_b$~\cite{Dowdall:2019bea},
\begin{equation}\label{eq:DmB_master_short}
\Delta m_{B_q}^{\phi}\simeq 2|M_{12}^{\phi}(B_q)|
=
\lambda_{q3}^2\,\overline{\Delta}_{B_q}(m_\phi)\,
\left[\frac{|R_2^{(q)}+R_4^{(q)}|}{4}\,f_{B_q}^2\,m_{B_q}\right],
\qquad
\overline{\Delta}_{B_q}(m_\phi)\equiv \frac{1}{q_{\rm eff}(B_q)^2+m_\phi^2},
\end{equation}
with $q=d,s$.

We use $q_{\rm eff}(B_d)=0.4226~{\rm GeV}$ and $q_{\rm eff}(B_s)=0.4412~{\rm GeV}$ from Ref.~\cite{Choi:2015ywa}. We take
$R_2^{(d)}=-2.059$, $R_4^{(d)}=3.822$, $R_2^{(s)}=-2.180$, $R_4^{(s)}=3.653$ from HPQCD~\cite{Dowdall:2019bea},
$f_{B_d}=189.5~{\rm MeV}$ and $f_{B_s}=230.3~{\rm MeV}$ from FLAG~\cite{FlavourLatticeAveragingGroupFLAG:2024oxs},
and $m_{B_d}=5.2796~{\rm GeV}$, $m_{B_s}=5.3669~{\rm GeV}$, $\Delta m_d=0.5069~{\rm ps}^{-1}$, and $\Delta m_s=17.765~{\rm ps}^{-1}$ from the PDG~\cite{ParticleDataGroup:2024cfk}.
In the ultralight regime $m_\phi\ll q_{\rm eff}(B_q)$, imposing $|\Delta m_{B_q}^{\phi}|\le (\Delta m_{B_q})_{\rm exp}$ yields
\begin{equation}
\lambda_{13}\lesssim 8.4\times 10^{-7},
\qquad
\lambda_{23}\lesssim 4.7\times 10^{-6}.
\label{eq:lambda13_23_bounds}
\end{equation}

\subsection{Exotic Force Measurements}
As described in Eq.~\eqref{eq:fifth-force}, our model mediates a composition--dependent ``fifth'' force: the exotic acceleration it induces on a test body depends on the atomic species. Precision tests of the weak equivalence principle (WEP) that compare the free-fall accelerations of different materials therefore provide direct constraints on this interaction.

A particularly stringent bound comes from the MICROSCOPE satellite mission, which operates on a nearly circular Earth orbit with radius
$R_{\rm MS}\approx 7000~\mathrm{km}$~\cite{MICROSCOPE:2022doy}. MICROSCOPE compares the accelerations of platinum and titanium test masses and reports a limit on the E\"otv\"os parameter:
\begin{equation}
\label{eq:eta_def}
\eta \equiv \frac{2|a_1-a_2|}{a_1+a_2}
\leqslant 2.7\times 10^{-15},
\end{equation}
where $a_i$ is the orbital free--fall acceleration of the $i$-th material. The common (composition--independent) part of the acceleration is gravitational,
\begin{equation}
\frac{a_1+a_2}{2}=\frac{G M_{\mathrm{E}}}{R_{\mathrm{MS}}^{2}},
\end{equation}
with $M_{\rm E}$ the Earth mass and $G$ the gravitational constant~\cite{ParticleDataGroup:2024cfk}. Any nonzero $\eta$ would then be sourced by the composition dependence of the exotic force in Eq.~\eqref{eq:fifth-force}.

In our setup, however, the effective coupling to nucleons is loop--generated, and the resulting differential acceleration is therefore parametrically suppressed. Translating the MICROSCOPE limit in Eq.~\eqref{eq:eta_def} into a bound on the nucleon effective scale, one finds that the constraint on the reciprocal coupling is weak,
\begin{equation}
\frac{1}{\Lambda_{p/n}} \lesssim \mathcal{O}(1)~\mathrm{GeV}^{-1},
\end{equation}
which implies a very mild restriction on $\lambda_{ij}$, much weaker than the bounds derived in the previous sections. We therefore do not display this constraint in the final plots.

\section{Astrophysical Constraints on ULDM}\label{sec:others}
One potential concern for our model is whether it is subject to astrophysical or stellar-cooling constraints, which commonly arise in searches for ultralight particles~\cite{Raffelt:1996wa,Heurtier:2016otg,Jaber-Urquiza:2023sal}. However, in the present setup---where the scalar field $\phi$ couples to quarks exclusively through flavor-off-diagonal interactions---such limits are either inapplicable or strongly suppressed.

Cooling processes in stars, red giants~\cite{Salaris_2018}, white dwarfs~\cite{Saumon:2022gtu}, or supernovae~\cite{Kamiokande-II:1987idp} take place in low-energy thermal plasmas ($T\sim {\rm keV}$-$\mathcal{O}(10)~{\rm MeV}$) whose relevant degrees of freedom are electrons, nucleons, and nuclei, i.e.\ matter that is effectively flavor-diagonal at hadronic scales. In this environment, the interaction $\phi\,\bar d_i d_j$ with $i\neq j$ corresponds, at the hadronic level, to transitions that change strangeness or bottomness. The emission of a single $\phi$ therefore requires either (i) external strange/bottom degrees of freedom in the medium, or (ii) the production of strange/bottom hadrons (or hypernuclear excitations) in the final state. Both possibilities are strongly disfavored in ordinary stellar interiors: there is no thermal population of strange or bottom hadrons, and the relevant hadronic thresholds (set by the masses of strange/bottom hadrons or the corresponding excitation energies) are generically $\mathcal{O}(100~{\rm MeV})$ or larger, far above the available thermal energies. As a result, flavor-changing emission channels such as $d\to s+\phi$ or $s\to b+\phi$ are either kinematically inaccessible or exponentially Boltzmann suppressed, rendering $\phi$ production through these off-diagonal couplings negligible for standard stellar-cooling considerations.

In dense neutron-star cores, $\phi$ emission could in principle occur if strange degrees of freedom---such as hyperons or kaon condensation---are present, in which case the medium is no longer purely flavor-diagonal. However, the existence and abundance of such states depend sensitively on the (still poorly constrained) equation of state of dense matter. Moreover, any coupling of $\phi$ to strange quarks relevant for these environments is already stringently constrained by terrestrial flavor experiments. In particular, rare kaon decays such as $K\to \pi+\phi$ directly probe the same flavor-changing interaction and provide robust, model-independent bounds that are typically stronger and far less uncertain than prospective neutron-star cooling limits.

Consequently, the usual stellar-cooling mechanisms are absent unless the model also contains flavor-diagonal components of the scalar coupling. Hence, for a scalar that couples exclusively through off-diagonal down-type quark interactions, astrophysical cooling constraints are subdominant, fragile, and less controllable than laboratory flavor-physics bounds.

\section{Conclusions}\label{sec:concl}
This work presents a systematic study of a flavor-violating ULDM portal in which a real scalar field $\phi$ couples off-diagonally to down-type quarks. We analyzed the same interaction in both regimes relevant to ultralight dark matter: a coherent classical background and a propagating quantum particle. In the coherent regime, the dominant effects are oscillatory shifts of down-type quark masses and CKM elements; in the particle regime, the same couplings contribute to flavor-violating decays and meson oscillations.

Using direct CKM inputs, time-resolved $^{37}\mathrm{K}$ and tritium data, quark-mass constraints, atomic clocks, pulsar timing, and meson mixing, we derived complementary bounds on $\lambda_{ij}$ across the mass range considered. In the coherent-field regime, clock and PTA observables provide the leading sensitivity, at about $\lambda_{12}\sim10^{-20}$, $\lambda_{13}\sim10^{-19}$, and $\lambda_{23}\sim10^{-18}$ in the ultralight region. Importantly, the $^{37}\mathrm{K}$ time-series analysis offers dedicated mid-frequency reach, yielding $\lambda_{12}\lesssim10^{-5}$ and $\lambda_{13}\lesssim10^{-3}$ around $m_\phi\sim10^{-14}\,\mathrm{eV}$.

In the particle interpretation, invisible meson decays and meson-mixing observables provide additional constraints, including $\lambda_{12}<2.1\times10^{-13}$ from $K\to\pi+\mathrm{Inv}$, $\lambda_{23}<2.1\times10^{-8}$ from $B\to K+\mathrm{Inv}$, and $\lambda_{13}<1.8\times10^{-8}$ from $B\to \pi+\mathrm{Inv}$, while meson oscillations provide further independent coverage of the parameter space.  Loop-induced fifth-force effects remain subdominant. Overall, combining coherent-field and particle probes significantly narrows flavor-violating ULDM parameter space over $m_\phi\sim10^{-24}\text{--}10^{-12}\,\mathrm{eV}$ and motivates future time-domain CKM analyses and dedicated flavor-violating decay searches.

\section{Acknowledgments}
The work of J.G is supported by the Postdoctoral Fellowship Program (Grade C) of China Postdoctoral Science Foundation under Grant No. GZC20252775.
The work of J.L. is supported by the National Science Foundation of China under Grant No. 12235001, No. 12475103
and State Key Laboratory of Nuclear Physics and Technology under Grant No. NPT2025ZX11. 
The work of X.P.W. is supported by National Science Foundation of China under Grant No. 12375095, and the Fundamental Research Funds for the Central Universities.
J.L. and X.P.W. thank the Asia Pacific Center for Theoretical Physics (APCTP), Pohang, Korea, for their hospitality during the focus program [APCTP-2025-F01], from which this work greatly benefited. J.L. and X.P.W. also thank the Mainz Institute for Theoretical Physics (MITP) of the PRISMA+ Cluster of Excellence (Project ID 390831469) for its hospitality and partial support during the completion of this work. We also acknowledge with appreciation the valuable discussions and insights provided by the members of the Collaboration of Precision Testing and New Physics.

\bibliographystyle{JHEP}
\bibliography{ref}

\end{document}